\documentclass[11pt,english,review]{elsarticle}
\usepackage[T1]{fontenc}
\usepackage[latin9]{inputenc}
\usepackage{geometry}
\usepackage{array}
\usepackage{bm}
\usepackage{multirow}
\usepackage{amsmath}
\usepackage{amssymb}
\usepackage{graphicx}
\usepackage{setspace}
\usepackage{esint}
\makeatletter

\providecommand{\tabularnewline}{\\}

\numberwithin{equation}{section}
\numberwithin{figure}{section}

\usepackage{lmodern}

\@ifundefined{showcaptionsetup}{}{%
 \PassOptionsToPackage{caption=false}{subfig}}
\usepackage{subfig}
\makeatother

\usepackage{babel}
\begin{document}

\begin{frontmatter}{}

\title{Distributed Detection of a Non-cooperative Target\\
via Generalized Locally-optimum Approaches }

\author[UNINA]{D. Ciuonzo\corref{mycorrespondingauthor}}

\cortext[mycorrespondingauthor]{Corresponding Author.}

\ead{domenico.ciuonzo@ieee.org}

\author[NTNU]{P. Salvo Rossi}

\ead{salvorossi@ieee.org}

\address[UNINA]{University of Naples \textquotedbl{}Federico II\textquotedbl{}, DIETI, Via Claudio 21, 80125 Naples, Italy.}

\address[NTNU]{Department of Electronics and Telecommunications, Norwegian University of Science and Technology, Trondheim, Norway.}
\begin{abstract}
In this paper we tackle distributed detection of a non-cooperative
target with a Wireless Sensor Network (WSN). When the target is present,
sensors observe an unknown random signal with amplitude attenuation
depending on the distance between the sensor and the target (unknown)
positions, embedded in white Gaussian noise. The Fusion Center (FC)
receives sensors decisions through error-prone Binary Symmetric Channels
(BSCs) and is in charge of performing a (potentially) more-accurate
global decision. The resulting problem is a one-sided testing with
nuisance parameters present only under the target-present hypothesis.
We first focus on fusion rules based on Generalized Likelihood Ratio
Test (GLRT), Bayesian and hybrid approaches. Then, aimed at reducing
the computational complexity, we develop fusion rules based on generalizations
of the well-known Locally-Optimum Detection (LOD) framework. Finally,
all the proposed rules are compared in terms of performance and complexity.\end{abstract}

\end{frontmatter}{}

\section{Introduction}

\subsection{Motivation and Related Works}

Wireless sensor networks (WSNs) have attracted significant attention
due to their potential in providing improved capabilities in performing
detection and estimation \cite{Ding2012,Ding2015}, reconnaissance
and surveillance, with a wide range of applications, comprising battlefield
surveillance, security, traffic, and environmental monitoring \cite{Chong2003}.
Distributed detection is among the fundamental tasks that a WSN needs
to accomplish which has been investigated in the recent years \cite{Varshney1996}.

Due to bandwidth and energy constraints, it is often assumed that
each sensor quantizes its own observation with a single bit before
transmission to the FC. This may be the result of a dumb quantization
\cite{Ciuonzo2013b,Fang2013} or represent the estimated decision
regarding the detection event \cite{Ciuonzo2014a,Ciuonzo2015a,P.SalvoRossi2015,P.SalvoRossi2015a}.
In the latter case, the decisions of individual sensors are collected
by the FC and combined according to a specifically-designed fusion
rule aiming at improved detection performance. In \cite{Chair1986},
the optimum strategy to fuse the local decisions at the FC has been
obtained under the conditional independence assumption. The optimal
fusion rule in both Neyman-Pearson and Bayesian senses, which is derived
from the likelihood ratio test \cite{Kay1998}, is commonly referred
to as \emph{Chair-Varshney} (CV) rule. It amounts to a threshold detector
on the weighted sum of binary sensor detections, with each weight
depending on sensor detection and false alarm probabilities.

\begin{figure}
\centering{}\includegraphics[width=0.7\paperwidth]{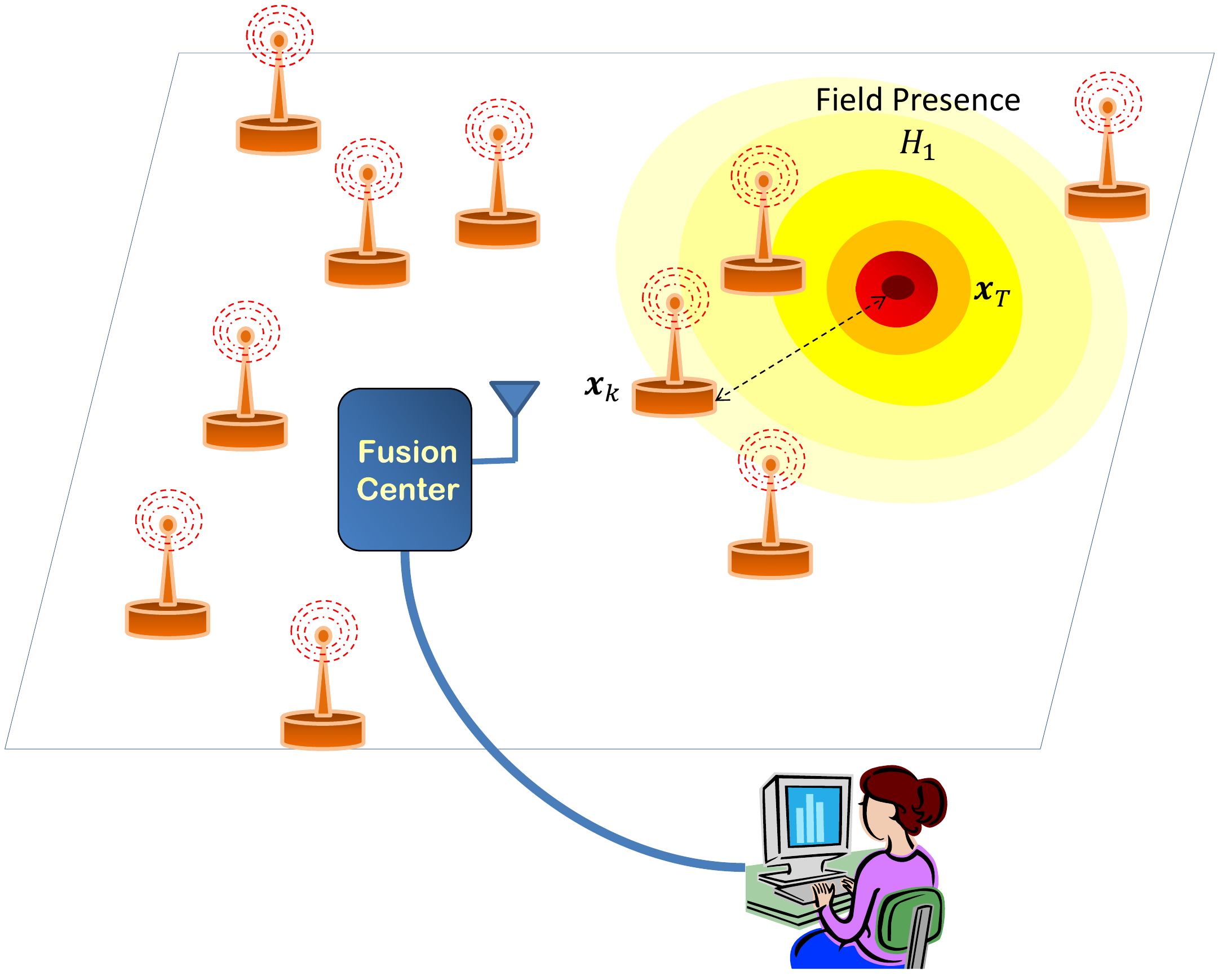}\caption{Distributed detection of a non-cooperative target with spatial signature:
System model. \label{fig: System Model}}
\end{figure}

Unfortunately, the local detection probability is seldom known or
difficult to estimate when the detection event relates to revealing
a target described by a spatial signature. In fact, in the latter
case the detection probability depends on the (unknown) constitutive
parameters of the target to be detected, such as the average power
and the target location (see Fig. \ref{fig: System Model}). Without
the knowledge of the local detection probabilities, the optimal fusion
rule becomes impractical and an attractive alternative is the so-called\emph{
Counting Rule} (CR) test, i.e. the FC counts the number of local detections
in the WSN and compares it with a threshold \cite{Niu2006}. A performance
analysis of the CR has been provided in \cite{Niu2008} for a WSN
with randomly deployed sensors. Unfortunately, CR suffers from performance
degradation when trying to detect spatial events. Indeed, though CR
is a very reasonable approach arising from different rationales \cite{Varshney1996,Ciuonzo2014a,Ciuonzo2015a},
it does not make any attempt to use information about the contiguity
of sensors that declare (potential) target presence. Therefore, based
on these considerations, several studies have focused on design of
fusion rules filling the performance gap between the CV rule and the
CR.

In \cite{Katenka2008} a two-step decision-fusion algorithm is proposed,
in which sensors first correct their decisions on the basis of neighboring
sensors, and then make a collective decision as a network. It is shown
that in many situations relevant to random sensor field detection,
the local vote correction achieves significantly higher target detection
probability than decision fusion based on the CR. Also, for the proposed
approach, an explicit formula for FC threshold choice (viz. false-alarm
rate determination) was provided, based on normal approximation of
the statistic under the target-absent hypothesis. A simple and more
accurate alternative for threshold choice based on the beta-binomial
approximation is proposed in \cite{Ridout2013}. In \cite{Niu2006b}
the Generalized Likelihood Ratio Test (GLRT) for the distributed detection
of a target with a deterministic Amplitude Attenuation Function (AAF)
and known emitted power is developed, and its superiority is shown
in comparison to the CR. It is worth noticing that a similar model
assuming a deterministic AAF was employed to analyze the (approximate)
theoretical performance of CR in \cite{Niu2008}. Differently, a stochastic
AAF (subsuming the Rayleigh fading model) is assumed in \cite{Guerriero2009}
and \cite{Guerriero2010}, the latter being able to account for possible
amplitude fluctuations. In the same works, also a scan statistic and
Bayesian-originated approaches were obtained and compared with existing
alternatives. In both works, the average emitted power of the target
is however assumed \emph{known}.

However, in many cases it is of practical importance to assume that
also the (average) target emitted power is not available at the FC,
which well fits the case of an \emph{uncooperative} target, i.e. there
is no preliminary agreement between target and sensors in order to
exchange the information related to the (average) emitted power or
make it possible to be estimated. Examples of practical interest for
an uncooperative target are the primary user in a cognitive-radio
system or an oil-spill source measured by an underwater sensor network.
To the best of authors' knowledge, a few works have dealt with the
latter case. In \cite{Shoari2012}, a GLRT was derived for the case
of unknown target position and emitted power and compared to the CR,
the CV rule and a GLRT based on the awareness of target emitted power.
It has been shown that the loss incurred by the proposed GLRT is marginal
when compared to the ``power-clairvoyant'' GLRT. Differently, in
\cite{Sung2005} an asymptotic locally-optimum detector was obtained
for a WSN with (random) sensors positions following a Poisson point
process. Remarkably, the aforementioned study accounted for unknown
emitted power. Unfortunately, the deterministic AAF there employed
implicitly assumed that FC has available the target position, thus
limiting its applicability, though some numerical analysis to investigate
mismatched AAF performance was provided. 

\subsection{Summary of Contributions}

In this paper, we focus on decentralized detection of a non-cooperative
target with a spatially-dependent emission (signature). We consider
the practical setup in which the received signal at each individual
sensor is embedded in white Gaussian noise\footnote{The Gaussian assumption for measurement noise is only made here for
the sake of simplicity; generalization of the present framework to
non-Gaussian noise is possible and will be object of future studies.} and affected by Rayleigh fading, with an AAF depending on the sensor-target
distance (viz. stochastic AAF). The Rayleigh fading assumption is
employed here to account for fluctuations of the transmitted signal
due to multipath propagation. For energy- and bandwidth-efficiency
purposes, each sensor performs a local decision on the absence/presence
of the target and forwards it to a FC, which is in charge of providing
a more accurate global decision. With reference to this setup, the
main contributions of the present work can be summarized as follows:
\begin{itemize}
\item We first review the scenario where the emitted power is available
(thus the sole target position is unknown) at the FC, in order to
understand the basics of the problem under investigation and list
various alternatives employed in the open literature, such as GLRT
\cite{Niu2006b} and Bayesian approaches. Then we switch to the more
realistic case of unknown target location and power, which is typical
in surveillance tasks. In this context we provide a systematic analysis
of several detectors based on: ($i)$ GLRT \cite{Shoari2012}, ($ii$)
Bayesian approach and ($iii$) hybrid combinations of the two (for
sake of completeness). 
\item In order to reduce the computational complexity required by these
approaches, we also develop two \emph{novel} sub-optimal fusion rules
based on the locally-optimum detection framework \cite{Kassam1988}.
The first relies on Bayesian assumption for the sole target position,
whereas the latter obviates the problem by resorting to Davies rationale
\cite{Davies1987}. The design and the analysis of such practical
rules and their comparison to the aforementioned alternatives represents
the main contribution of this work. We underline that, since a uniformly
most powerful test does not exist for our problem (because of the
unknown parameters), nothing can be said in advance on their relative
performance. All the aforementioned detectors are also compared in
terms of computational complexity;
\item The scenario at hand is then extended to the demanding case of imperfect
reporting channels (typical for battery-powered sensors implementing
low-energy communications), modeled as Binary Symmetric Channels (BSCs).
The proposed fusion rules are then extended to take into account the
(additional) reporting uncertainty, under the assumption of \emph{known}
Bit-Error Probabilities (BEPs).
\item Finally, simulation results are provided to compare all the considered
rules in some practical scenarios and to underline the relevant trends.
\end{itemize}

\subsection{Paper Organization and Manuscript Notation}

The remainder of the paper is organized as follows: in Sec.~\ref{sec: System model}
we describe the system model, with reference to local sensing and
FC modeling. In Sec.~\ref{sec: Known target power} we recall and discuss
the problem of distributed detection under the assumption of a known
average target emitted power. Differently, Sec.~\ref{sec: Unknown target power}
is devoted to the development of fusion rules which deal with the
additional uncertainty of unknown power. Then, in Sec.~\ref{sec: BSC reporting channels}
we extend the obtained fusion rules to the more general case of imperfect
reporting channels between the sensors and the FC. All the considered
rules are compared in terms of complexity in Sec.~\ref{sub: Complexity comparison}.
Furthermore, in Sec.~\ref{sec: Simulation results} a set of simulations
is provided to compare the developed rules and assess the loss incurred
by non-availability of emitted power. Finally, some conclusions are
drawn in Sec.~\ref{sec: Conclusions}. Proofs and derivations are
confined to a dedicated Appendix.

\emph{Notation} - Lower-case bold letters denote vectors, with $a_{n}$
being the $n$th element of $\bm{a}$; upper-case calligraphic letters,
e.g. $\mathcal{A}$, denote finite sets; $\mathbb{E}\{\cdot\}$, $\mathrm{var\{\cdot\}}$,
$(\cdot)^{T}$, and $\left\Vert \cdot\right\Vert $ denote expectation,
variance, transpose and Euclidean norm operators, respectively; $P(\cdot)$
and $p(\cdot)$ denote probability mass functions (pmfs) and probability
density functions (pdfs), while $P(\cdot|\cdot)$ and $p(\cdot|\cdot)$
their corresponding conditional counterparts; $\mathcal{N}(\mu,\sigma^{2})$
denotes a Gaussian pdf with mean $\mu$ and variance $\sigma^{2}$;
$\mathcal{Q}(\cdot)$ is the complementary cumulative distribution
function (ccdf) of a standard normal random variable; finally the
symbols $\propto$ and $\sim$ mean ``statistically equivalent to''
and \textquotedblleft distributed as\textquotedblright , respectively.

\section{System Model\label{sec: System model}}

We consider a scenario where $K$ sensors are deployed in a surveillance
area to monitor the absence ($\mathcal{H}_{0}$) or presence ($\mathcal{H}_{1}$)
of a target of interest having a spatial signature. The measurement
model of the generic sensor is described in Sec. \ref{sub: Sensing Model}.
Then, we introduce the local decision procedure employed (independently)
by each sensor in Sec. \ref{sub: Local decision model}. Finally,
in Sec. \ref{sub: Decision Fusion} we describe the problem of fusing sensors
decisions at the FC.

\subsection{Sensing Model\label{sub: Sensing Model}}

When the target is present in the surveillance area (i.e. $\mathcal{H}_{1}$),
we assume that its radiated signal is isotropic and experiences (distance-depending)
path-loss, Rayleigh fading, and Additive White Gaussian Noise (AWGN),
before reaching individual sensors. In other terms, the sensing model
for $k$th sensor ($k\in\{1,\ldots,K\}$) under $\mathcal{H}_{1}$
is \cite{Guerriero2010} 
\begin{equation}
y_{k}=\xi_{k}\,g(\bm{x}_{T},\bm{x}_{k})+w_{k}\,,\label{eq: Sensing model}
\end{equation}
where $y_{k}\in\mathbb{R}$ is signal measured by $k$th sensor and
$w_{k}\sim\mathcal{N}(0,\sigma_{w,k}^{2})$ denotes the corresponding
measurement noise. Furthermore, $\bm{x}_{T}\in\mathbb{R}^{d}$ denotes
the\emph{ unknown} position of the target (in $d$-dimensional coordinates),
while $\bm{x}_{k}\in\mathbb{R}^{d}$ denotes the \emph{known} $k$th
sensor position (in $d$-dimensional coordinates). The positions $\bm{x}_{T}$
and $\bm{x}_{k}$ uniquely determine the value of $g(\bm{x}_{T},\bm{x}_{k})$,
generically denoting the AAF. Finally $\xi_{k}$ is a Gaussian distributed
random variable, $\xi_{k}\sim\mathcal{N}(0,\sigma_{s}^{2})$, modelling
fluctuations in the received signal strength at $k$th sensor. Due
to spatial separation of the sensors, we assume that the noise contributions
$w_{k}$s and the fading coefficients $\xi_{k}$s are both statistically
independent. Depending on the peculiar scenario being investigated,
$\sigma_{s}^{2}$ will be assumed either \emph{known }(Sec. \ref{sec: Known target power})
or \emph{unknown} (Sec. \ref{sec: Unknown target power}).

Then, the measured signal $y_{k}$ is distributed under hypotheses
$\mathcal{H}_{0}$ and $\mathcal{H}_{1}$ as
\begin{equation}
y_{k}\,|\,\mathcal{H}_{0}\sim\mathcal{N}(0,\sigma_{w,k}^{2}),\qquad\quad y_{k}\,|\,\mathcal{H}_{1}\sim\mathcal{N}\left(0,\,\sigma_{s}^{2}\,g^{2}(\bm{x}_{T},\bm{x}_{k})+\sigma_{w,k}^{2}\right),
\end{equation}
respectively. With reference to the specific AAF, two common examples
\cite{Katenka2008,Guerriero2010} are the \emph{power-law} attenuated
model
\begin{equation}
g(\bm{x}_{T},\bm{x}_{k})\triangleq\frac{1}{\sqrt{1+\left(\frac{\left\Vert \bm{x}_{T}-\bm{x}_{k}\right\Vert }{\eta}\right)^{\alpha}}}\,,\label{eq: power-law attenuation function}
\end{equation}
and the \emph{exponentially} attenuated model 
\begin{equation}
g(\bm{x}_{T},\bm{x}_{k})\triangleq\sqrt{\exp\left(-\frac{\left\Vert \bm{x}_{T}-\bm{x}_{k}\right\Vert ^{2}}{\eta^{2}}\right)}\,.\label{eq: exponentially-attenuated power}
\end{equation}
In Eqs. (\ref{eq: power-law attenuation function}) and (\ref{eq: exponentially-attenuated power})
the parameter $\eta$ controls the (approximate) spatial signature
extent produced by both AAFs, while $\alpha$ is a positive coefficient
that dictates the rapidity of signal decay as a function of the distance
in the case of power-law AAF.

\begin{figure}
\begin{centering}
\includegraphics[width=0.9\paperwidth]{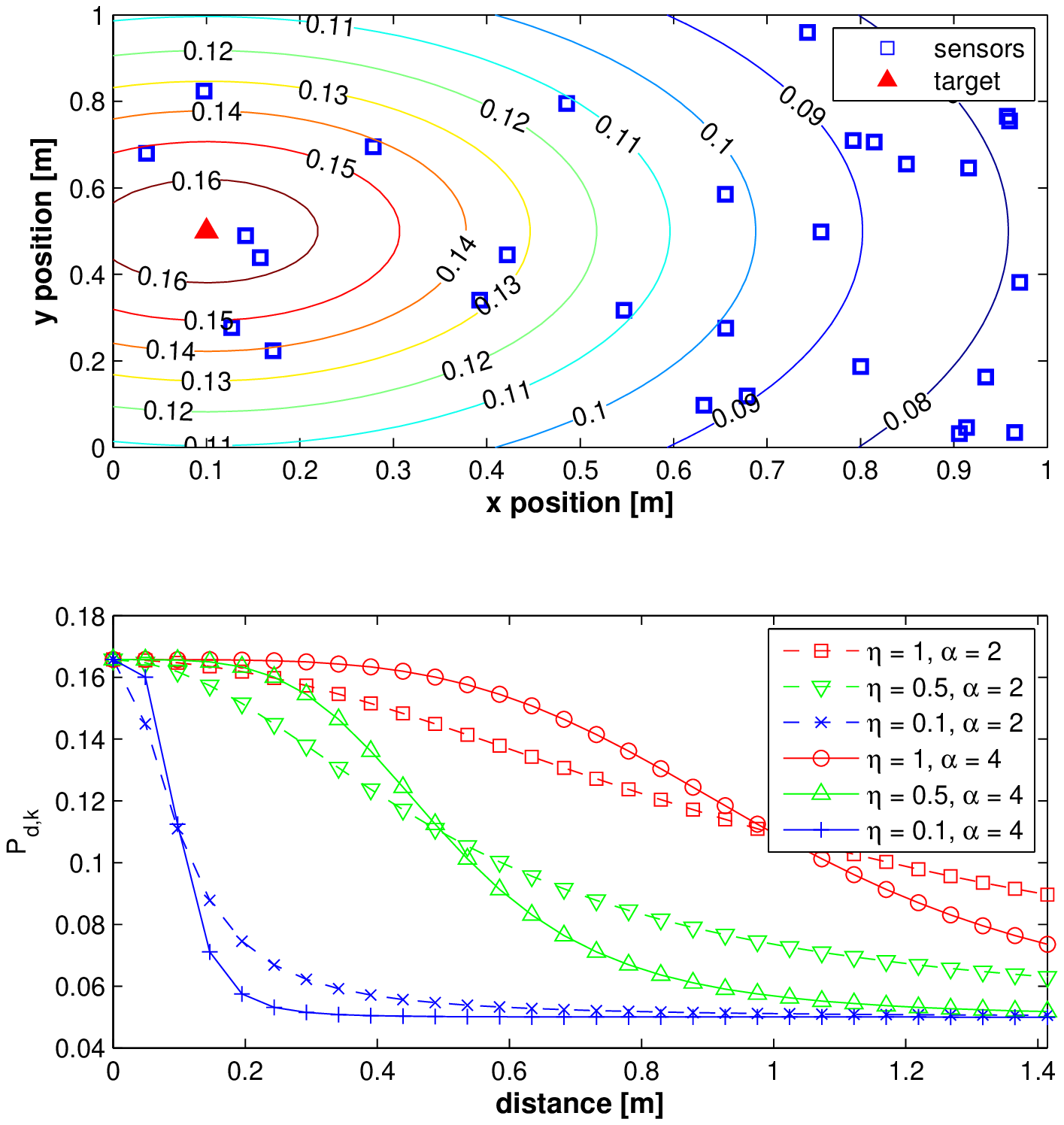}
\par\end{centering}

\centering{}\caption{Detection probability ($P_{d,k}$) field, for a fixed $P_{f,k}=0.05$:
Power-law AAF. Top plot shows $P_{d,k}$ vs. $\bm{x}$ (generic sensor
position) for a target located at $\bm{x}_{T}=[0.1\;0.5]^{T}$; bottom
plot depicts $P_{d,k}$ vs. $\left\Vert \bm{x}_{T}-\bm{x}\right\Vert $.
\label{fig: Det_prob_field (plaw)}}
\end{figure}
\begin{figure}
\begin{centering}
\includegraphics[width=0.9\paperwidth]{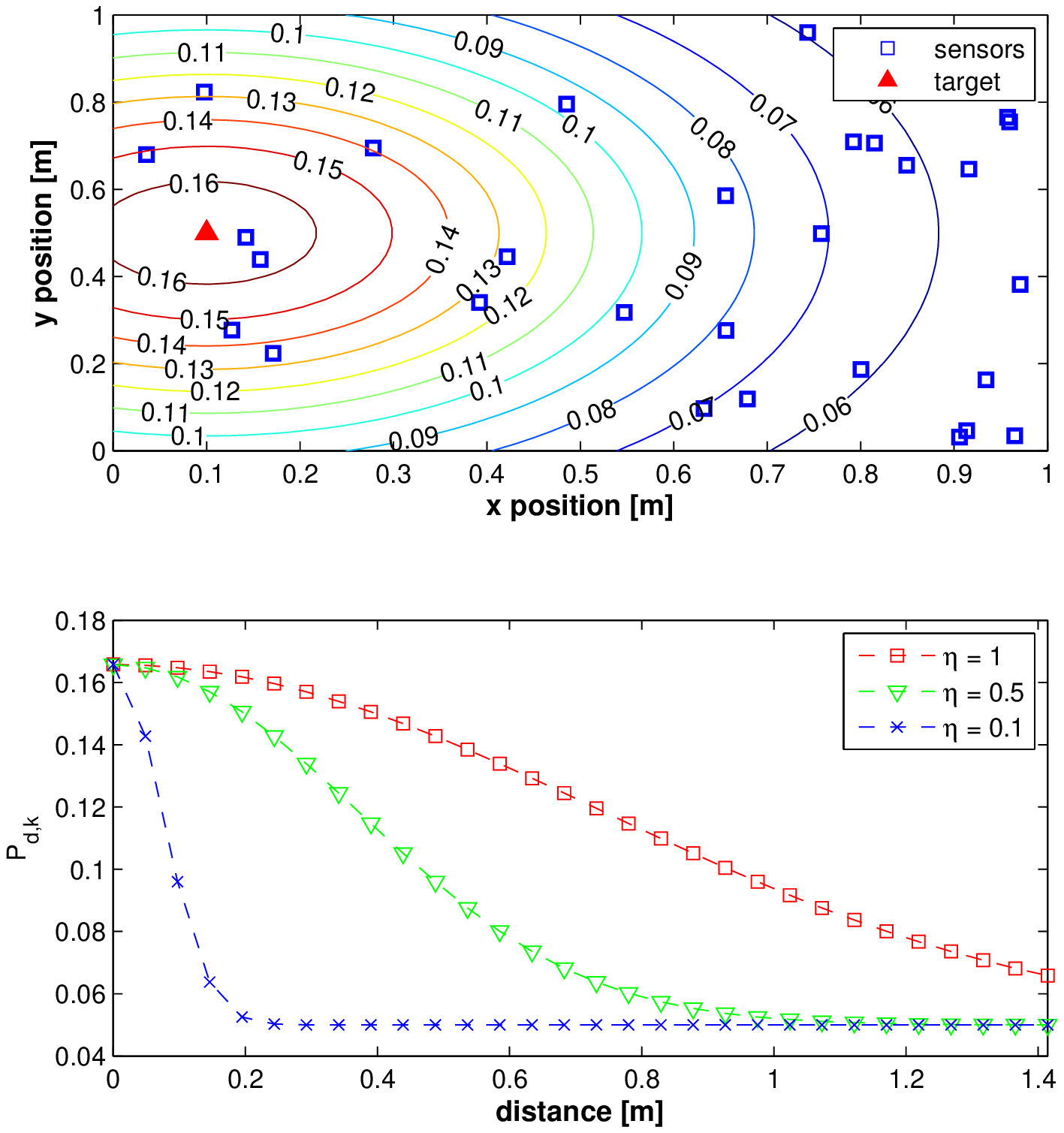}
\par\end{centering}

\centering{}\caption{Detection probability ($P_{d,k}$) field, for a fixed $P_{f,k}=0.05$:
Exponential AAF. Top plot shows $P_{d,k}$ vs. $\bm{x}$ (generic
sensor position) for a target located at $\bm{x}_{T}=[0.1\;0.5]^{T}$;
bottom plot depicts $P_{d,k}$ vs. $\left\Vert \bm{x}_{T}-\bm{x}\right\Vert $.\label{fig: Det_prob_field (exp)}}
\end{figure}

\subsection{Local Decision Approach\label{sub: Local decision model}}

We assume that sensors make their local decisions \emph{individually}
without collaboration. Then, each sensor is faced to tackle the following
\emph{composite} hypothesis testing:
\begin{equation}
\begin{cases}
\mathcal{H}_{0}\,:\, & y_{k}=w_{k}\\
\mathcal{H}_{1}\,:\, & y_{k}=\xi_{k}\,g(\bm{x}_{T},\bm{x}_{k})+w_{k}
\end{cases}\label{eq: local sensor hypothesis testing problem}
\end{equation}
Indeed, although the sensor may be aware of its own position $\bm{x}_{k}$,
the target position $\bm{x}_{T}$ is clearly \emph{unknown}, independently
from the availability of the average emitted power $\sigma_{s}^{2}$.
Nonetheless, for this specific sensing model it can be shown that
this difficulty can be elegantly circumvented. To this end, we consider
a local decision procedure based on the well-known Neyman-Pearson
lemma \cite{Kay1998}. More specifically, we consider the \emph{local}
Log Likelihood-Ratio (LLR) of $k$th sensor, denoted with $\lambda_{k}$,
whose explicit expression is: 
\begin{align}
\lambda_{k} & \triangleq\ln\left[\frac{p(y_{k}|\mathcal{H}_{1})}{p(y_{k}|\mathcal{H}_{0})}\right]\nonumber \\
 & =\frac{1}{2}\ln\left[\frac{\sigma_{w,k}^{2}}{\sigma_{w,k}^{2}+\sigma_{s}^{2}\,g^{2}(\bm{x}_{T},\bm{x}_{k})}\right]+\frac{\sigma_{s}^{2}\,g^{2}(\bm{x}_{T},\bm{x}_{k})}{\sigma_{w,k}^{2}\left[\sigma_{w,k}^{2}+\sigma_{s}^{2}\,g^{2}(\bm{x}_{T},\bm{x}_{k})\right]}\,y_{k}^{2}\,,\label{eq: local LLR kth sensor}
\end{align}
Last equation reveals that the LLR is an increasing function of $y_{k}^{2}$,
irrespective of the target average emitted power $\sigma_{s}^{2}$
and the target location $\bm{x}_{T}$. Therefore, by Karlin-Rubin
theorem \cite{Lehmann2006}, the following energy test
\begin{equation}
y_{k}^{2}\begin{array}{c}
{\scriptstyle \hat{\mathcal{H}}=\mathcal{H}}_{1}\\
\gtrless\\
{\scriptstyle \hat{\mathcal{H}}=\mathcal{H}}_{0}
\end{array}\gamma_{k}\,\label{eq: UMP local test}
\end{equation}
is \emph{Uniformly Most Powerful}\footnote{We recall that in the case of composite hypothesis testing problems
the uniformly most powerful test seldom exists \cite{Kay1998}. In
the latter case, alternative approaches such as the GLRT may be pursued.} (UMP)\emph{ }in a local sense. Also, $\gamma_{k}$ is a suitable
threshold chosen to ensure a certain false-alarm rate at the sensor
(in Neyman-Pearson approach) or to minimize the error-probability
(in the Bayesian framework). In view of the aforementioned considerations,
in what follows we will assume that each sensor implements its\emph{
local} UMP test based on its (local) measurement $y_{k}$. 

Furthermore, we observe that the performance of the energy test in
Eq. (\ref{eq: UMP local test}) is easily obtained explicitly, in
terms of the detection ($P_{d,k}\triangleq\Pr\{\lambda_{k}\geq\gamma_{k}|\mathcal{H}_{1}\}$)
and false-alarm ($P_{f,k}\triangleq\Pr\{\lambda_{k}\geq\gamma_{k}|\mathcal{H}_{0}\}$)
probabilities as \cite{Kay1998}
\begin{align}
P_{d,k} & =2\,\mathcal{Q}\left(\sqrt{\frac{\gamma_{k}}{\sigma_{w,k}^{2}+\sigma_{s}^{2}\,g^{2}(\bm{x}_{T},\bm{x}_{k})}}\right);\qquad P_{f,k}=2\,\mathcal{Q}\left(\sqrt{\frac{\gamma_{k}}{\sigma_{w,k}^{2}}}\right).\label{eq: Pd Pfa local sensors}
\end{align}
Two examples of a $P_{d,k}$ \emph{field} (that is, the detection
probability vs. the generic sensor position $\bm{x}$ for fixed target
position and false-alarm probability) are depicted in the top plots
of Figs. \ref{fig: Det_prob_field (plaw)} and \ref{fig: Det_prob_field (exp)},
for the power-law and exponential AAFs, respectively, with reference
to a 2-D square surveillance area of length $L=1$. Also, we assumed
$\bm{x}_{T}=\begin{bmatrix}0.1 & 0.5\end{bmatrix}^{T}$, $\eta=0.5$,
$\alpha=2$, $\sigma_{w,k}^{2}=1$, $\sigma_{s}^{2}=1$ and $P_{f,k}=0.05$
(from which $\gamma_{k}$ is easily deduced, cf. Eq.~(\ref{eq: Pd Pfa local sensors})).
Similarly, in the bottom plots of Figs. \ref{fig: Det_prob_field (plaw)}
and \ref{fig: Det_prob_field (exp)}, we have showed the same $P_{d,k}$
(with the same parameters as the top plots) as a function of the distance
$\left\Vert \bm{x}_{T}-\bm{x}\right\Vert $ for different values of
$\eta\in\{0.1,0.5,1\}$ and $\alpha\in\{2,4\}$ (in the case of power
law AAF).

Without loss of generality, we assume that $k$th sensor decision,
denoted as $d_{k}$, follows the map $d_{k}=i$ when hypothesis $\mathcal{H}_{i}$
is declared. Finally, for the sake of notational compactness, we define
the vector $\bm{d}\triangleq\begin{bmatrix}d_{1} & \cdots & d_{K}\end{bmatrix}^{T}$.

\subsection{Decision Fusion\label{sub: Decision Fusion}}

Each sensor then sends its decision $d_{k}$ to the FC, which employs
a threshold-based decision test (we interchangeably use the term ``fusion
rule'') on the basis of vector $\bm{d}$, that is:
\begin{equation}
\Lambda(\bm{d})\begin{array}{c}
{\scriptstyle \hat{\mathcal{H}}=\mathcal{H}}_{1}\\
\gtrless\\
{\scriptstyle \hat{\mathcal{H}}=\mathcal{H}}_{0}
\end{array}\bar{\gamma}\,,
\end{equation}
where $\bar{\gamma}$ is the threshold chosen to ensure a certain
\emph{global} false-alarm rate at the sensor (in Neyman-Pearson approach)
or to minimize the \emph{global} fusion error-probability (in the
Bayesian framework) \cite{Kay1998}. Global performance are accordingly
evaluated in terms of (global) probability of false alarm ($P_{f,0}$
) and detection ($P_{d,0}$), defined as follows
\begin{equation}
P_{f,0}\triangleq\Pr(\Lambda>\bar{\gamma}|\mathcal{H}_{0}),\qquad P_{d,0}\triangleq\Pr(\Lambda>\bar{\gamma}|\mathcal{H}_{1})\,.
\end{equation}
It is worth noticing that $\Pr(\Lambda>\bar{\gamma}|\mathcal{H}_{i})$
generically describes both $P_{f,0}$ and $P_{d,0}$ (with $i=0$
and $i=1$, respectively). The behavior of the global probability
of detection ($P_{d,0}$) versus the global probability of false alarm
($P_{f,0}$) is commonly denoted Receiver Operating Characteristic
(ROC) \cite{Varshney1996}.

It is apparent that, under hypothesis $\mathcal{H}_{1}$, the pmf
of $\bm{d}$ assumes the explicit expression represented by the product
of independent Bernoulli pmfs (since the decisions $d_{k}$ are conditionally
\emph{independent}, as an immediate consequence of mutual independence
of $w_{k}$s, $\xi_{k}$s and of decoupled quantization process),
that is
\begin{equation}
P(\bm{d}|\mathcal{H}_{1})=\prod_{k=1}^{K}P(d_{k}|\mathcal{H}_{1})=\prod_{k=1}^{K}(P_{d,k})^{d_{k}}(1-P_{d,k})^{(1-d_{k})}\,,
\end{equation}
and a similar expression holds for $P(\bm{d}|\mathcal{H}_{0})$, when
replacing $P_{d,k}$ with $P_{f,k}$. The optimal decision statistic
in both Neyman-Pearson and Bayesian senses is represented by the (global)
LLR, given by
\begin{align}
\Lambda_{{\scriptscriptstyle \mathrm{LLR}}} & \triangleq\ln\left[\frac{P(\bm{d}|\mathcal{H}_{1})}{P(\bm{d}|\mathcal{H}_{0})}\right]=\sum_{k=1}^{K}\ln\left[\frac{P(d_{k}|\mathcal{H}_{1})}{P(d_{k}|\mathcal{H}_{0})}\right]\nonumber \\
 & =\sum_{k=1}^{K}\left\{ d_{k}\ln\left[\frac{P_{d,k}}{P_{f,k}}\right]+(1-d_{k})\ln\left[\frac{1-P_{d,k}}{1-P_{f,k}}\right]\right\} \,,\label{eq:Clairvoyant_LR}
\end{align}
where $P_{d,k}$ and $P_{f,k}$ are defined in Eq. (\ref{eq: Pd Pfa local sensors}).
Unfortunately, the LLR \emph{cannot be implemented} as the $P_{d,k}$s
are usually \emph{unknown}, since they depend on the constitutive
parameters of the (unknown) target emission, that is ($i$) the average
power $\sigma_{s}^{2}$ and ($ii$) the target location $\bm{x}_{T}$.
Therefore, it is apparent that Eq.~(\ref{eq:Clairvoyant_LR}) should
not be intended as a realistic element of comparison, but rather as
an optimistic (upper) bound on the achievable performance, based on
a clairvoyant assumption.

On the other hand, from direct inspection of Eq. (\ref{eq: Pd Pfa local sensors}),
we notice that $P_{d,k}\geq P_{f,k}$ $\forall k\in\mathcal{K}$,
as each reasonable local decision procedure\footnote{In other terms, the applicability of CR to decision fusion is also
valid in the case of local decision statistics at the sensors based
on sub-optimal approaches, as opposed to what is assumed in this manuscript.} would achieve a ROC \textemdash{} operation point that is more informative
than an unbiased coin (i.e. above the chance line). Based on this
observation, we may apply the well-known \emph{Counting Rule} (CR)
\cite{Varshney1996}, not requiring sensors local performance for
its implementation. This rule is widely used in DF (due to its simplicity
and no requirements on system knowledge) and based on the following
statistic:
\begin{equation}
\Lambda_{{\scriptscriptstyle \mathrm{CR}}}\triangleq\sum_{k=1}^{K}d_{k}\,.\label{eq:Counting_rule}
\end{equation}
The above rule, despite of its simplicity, has been obtained under
different rationales in the literature \cite{Varshney1996,Ciuonzo2014a,Ciuonzo2015a}.
In fact, Eq. (\ref{eq:Counting_rule}) can be obtained as follows:
\begin{itemize}
\item It is statistically equivalent to the LLR in Eq. (\ref{eq:Clairvoyant_LR}),
by assuming that the sensors have all equal performance (i.e., $P_{d,k}=P_{d}$
and $P_{f,k}=P_{f}$) \cite{Varshney1996};
\item It is statistically equivalent to the locally most-mean powerful test
\cite{Gupta1986} in a partially-homogeneous scenario ($P_{f,k}=P_{f}$),
assuming no other constraint than $P_{d,k}\geq P_{f,k}$ \cite{Ciuonzo2015a};
\item It is the UMP \emph{invariant} test for the permutation group \cite{Ciuonzo2015a};
\item It is statistically equivalent to the GLRT and Rao test in a partially-homogeneous
scenario ($P_{f,k}=P_{f}$, when $P_{f}<\frac{1}{2}$), assuming no
other constraint than $P_{d,k}\geq P_{f,k}$ \cite{Ciuonzo2015a}.
\end{itemize}

\section{Practical Fusion Rules}

\subsection{Known Target Power\label{sec: Known target power}}

Initially, we assume that $\sigma_{s}^{2}$ is \emph{known} and then
the hypothesis testing problem can be summarized as:
\begin{equation}
\begin{cases}
\mathcal{H}_{0}\,:\, & \sigma_{s}^{2}=0\\
\mathcal{H}_{1}\,:\, & \sigma_{s}^{2}>0\,(\mathrm{known})\,,\quad\bm{x}_{T}\,(\mathrm{nuisance})
\end{cases},
\end{equation}
that is, we are concerned with discriminating between two hypotheses
where the nuisance parameters ($\bm{x}_{T}$) are present \emph{only}
under the (alternative) hypothesis $\mathcal{H}_{1}$. Such case has
been analyzed in the works \cite{Niu2006,Guerriero2010}.

\subsubsection*{GLRT}

In \cite{Niu2006b} the authors proposed the use of a GLR statistic\footnote{We recall that in general the GLRT requires the evaluation of the
MLE of the unknown parameter set under \emph{both} hypotheses. However,
referring to our specific case, the nuisance parameter $\bm{x}_{T}$
is \emph{not observable }under $\mathcal{H}_{0}$. Therefore, the
pdf of null hypothesis is completely specified and no MLE evaluation
is required in the latter case.}, whose explicit log form for the considered problem is
\begin{align}
\Lambda_{\mathrm{G}} & \triangleq\ln\left[\frac{\max_{\bm{x}_{T}}P(\bm{d}|\mathcal{H}_{1};\bm{x}_{T})}{P(\bm{d}|\mathcal{H}_{0})}\right]\\
 & =\sum_{k=1}^{K}\left\{ d_{k}\,\ln\left[\frac{P_{d,k}(\widehat{\bm{x}}_{T})}{P_{f,k}}\right]+(1-d_{k})\,\ln\left[\frac{1-P_{d,k}(\widehat{\bm{x}}_{T})}{1-P_{f,k}}\right]\right\} \,,\label{eq: GLRT}
\end{align}
where $\widehat{\bm{x}}_{T}$ denotes the Maximum Likelihood Estimate
(MLE) of the target position, assuming that $\mathcal{H}_{1}$ holds,
that is:
\begin{equation}
\widehat{\bm{x}}_{T}\triangleq\arg\max_{\bm{x}_{T}}\,P(\bm{d}|\mathcal{H}_{1};\bm{x}_{T})\,.\label{eq: MLE x_T known power}
\end{equation}
Clearly, the higher the estimation accuracy of $\bm{x}_{T}$, the
higher the performance of GLR statistic. It is worth noticing that
$\widehat{\bm{x}}_{T}$ cannot be obtained in closed form and therefore
a grid search (or optimization routines) should be devised (details
on implementation are later provided in Sec. \ref{sub: Complexity comparison}).
Exploiting the parametric independence of $P(\bm{d}|\mathcal{H}_{0})$
on $\bm{x}_{T}$ and the monotonic property of logarithm, the above
expression can be rewritten in terms of Eq. (\ref{eq:Clairvoyant_LR})
as:
\begin{equation}
\Lambda_{\mathrm{G}}=\max_{\bm{x}_{T}}\,\Lambda_{{\scriptscriptstyle \mathrm{LLR}}}(\bm{x}_{T})\,,\label{eq: GLR - known power scenario}
\end{equation}
where $\Lambda_{{\scriptscriptstyle \mathrm{LLR}}}(\bm{x}_{T})$ underlines
the evaluation of LLR in Eq. (\ref{eq:Clairvoyant_LR}) assuming that
the target position equals $\bm{x}_{T}$. The alternative form in
(\ref{eq: GLR - known power scenario}) will be exploited to draw
out interesting considerations when comparing GLRT with other detectors.

\subsubsection*{Bayesian Approach}

The pdf dependence on target\textquoteright s position under $\mathcal{H}_{1}$
may be eliminated if a prior distribution on the position itself is
available (or can be safely assumed) and integrating the corresponding
likelihood (see, for example, \cite{Berger1999} for the advantages
provided by the Bayesian approach). Then, the explicit expression
of the Bayesian LLR is given by \cite{Guerriero2010}
\begin{align}
\Lambda_{\mathrm{B}} & \triangleq\ln\left[\frac{\int P(\bm{d}|\mathcal{H}_{1};\bm{x}_{T})\,p(\bm{x}_{T})d\bm{x}_{T}}{P(\bm{d}|\mathcal{H}_{0})}\right]\nonumber \\
 & =\ln\int\prod_{k=1}^{K}\left(\frac{P_{d,k}(\bm{x}_{T})}{P_{f,k}}\right)^{d_{k}}\left(\frac{1-P_{d,k}(\bm{x}_{T})}{1-P_{f,k}}\right)^{(1-d_{k})}\,p(\bm{x}_{T})\,d\bm{x}_{T}\,,\label{eq: LR Bayesian target pos}
\end{align}
where the dependence of $P_{d,k}$ on target position $\bm{x}_{T}$
is underlined. It is interesting to notice that the above expression
can be rewritten as:
\begin{equation}
\Lambda_{\mathrm{B}}=\ln\int\exp\left(\Lambda_{{\scriptscriptstyle \mathrm{LLR}}}(\bm{x}_{T})\right)\,p(\bm{x}_{T})\,d\bm{x}_{T}\,,\label{eq: Bayesian approach and LLR correspondence}
\end{equation}
where $\Lambda_{{\scriptscriptstyle \mathrm{LLR}}}(\bm{x}_{T})$ has
an analogous definition as that in Eq. (\ref{eq: GLR - known power scenario}).

\subsection{Unknown Target Power\label{sec: Unknown target power}}

Differently, when $\sigma_{s}^{2}$ is assumed \emph{unknown,} the
resulting (composite) hypothesis testing generalizes to:
\begin{equation}
\begin{cases}
\mathcal{H}_{0}\,:\, & \sigma_{s}^{2}=0\\
\mathcal{H}_{1}\,:\, & \sigma_{s}^{2}>0\,,\quad\bm{x}_{T}\,(\mathrm{nuisance})
\end{cases}\label{eq: Hypothesis testing -  Unknown Target Power}
\end{equation}
The above problem is recognized as a one-sided hypothesis testing
with nuisance parameters that are present only under the (alternative)
hypothesis $\mathcal{H}_{1}$. In the rest of the paper, for the sake
of notational convenience, we will use the symbol $\theta$ to refer
to the unknown average power $\sigma_{s}^{2}$ (with corresponding
notation $\theta_{0}$ for $\sigma_{s}^{2}=0$).

\subsubsection*{Discussion: Counting Rule (CR) and Clairvoyant LLR}

It is worth noticing that, in the case of unknown power $\sigma_{s}^{2}$,
the CR can be still implemented, as it does not require the knowledge
of the $P_{d,k}$s (cf. Eq. (\ref{eq:Counting_rule})). Similarly,
in the present scenario we will refer to the statistic which has (unrealistic)
knowledge of both $\bm{x}_{T}$ and $\sigma_{s}^{2}$ as a clairvoyant
LLR and thus the same formula as in Eq. (\ref{eq:Clairvoyant_LR})
can be applied. Apparently, in the considered scenario, the LLR will
represent an even \emph{looser} benchmark on the performance of practical
fusion rules.

\subsubsection*{GLRT\label{sub: unknown power - fully GLRT}}

The GLRT for this case was proposed and analyzed in \cite{Shoari2012}.
Indeed, the explicit expression of the (log-)GLR statistic is:
\begin{align}
\Lambda_{\mathrm{G}} & \triangleq\ln\left[\frac{\max_{\sigma_{s}^{2},\bm{x}_{T}}P(\bm{d}|\mathcal{H}_{1};\bm{x}_{T},\sigma_{s}^{2})}{P(\bm{d}|\mathcal{H}_{0})}\right]\\
 & =\sum_{k=1}^{K}\left\{ d_{k}\,\ln\left[\frac{P_{d,k}\left(\widehat{\bm{x}}_{T},\widehat{\sigma_{s}^{2}}\right)}{P_{f,k}}\right]+(1-d_{k})\,\ln\left[\frac{1-P_{d,k}\left(\widehat{\bm{x}}_{T},\widehat{\sigma_{s}^{2}}\right)}{1-P_{f,k}}\right]\right\} \,,\label{eq: GLRT-unknown power}
\end{align}
where $\widehat{\bm{x}}_{T}$ and $\widehat{\sigma_{s}^{2}}$ denote
the ML estimates of the target position and (average) target power,
assuming that hypothesis $\mathcal{H}_{1}$ is true, that is:
\begin{equation}
\left(\widehat{\bm{x}}_{T},\widehat{\sigma_{s}^{2}}\right)\triangleq\arg\max_{\bm{x}_{T},\,\sigma_{s}^{2}}\,P(\bm{d}|\mathcal{H}_{1};\bm{x}_{T},\sigma_{s}^{2})\,.\label{eq: MLE x_T, sigma_s2}
\end{equation}
The above expression can be rewritten (similarly as in the case of
known power, cf. Eq. (\ref{eq: GLR - known power scenario})) in terms
of Eq. (\ref{eq:Clairvoyant_LR}) as:
\begin{equation}
\Lambda_{\mathrm{G}}=\max_{\bm{x}_{T},\sigma_{s}^{2}}\,\Lambda_{{\scriptscriptstyle \mathrm{LLR}}}(\bm{x}_{T},\sigma_{s}^{2})\,,\label{eq: GLR - Unknown power}
\end{equation}
where $\Lambda_{{\scriptscriptstyle \mathrm{LLR}}}(\bm{x}_{T},\sigma_{s}^{2})$
is used to denote the LLR of Eq. (\ref{eq:Clairvoyant_LR}) evaluated
assuming the target position and power corresponding to $\bm{x}_{T}$
and $\sigma_{s}^{2}$, respectively.

\subsubsection*{Bayesian Approach\label{sub: unknown power Fully Bayesian}}

In order to follow a purely Bayesian approach, we need eliminate both
the dependence on target\textquoteright s position and average emitted
power (under $\mathcal{H}_{1}$) by assigning prior distributions
to them both and integrating the corresponding likelihood. Thus, the
closed form of the Bayesian LLR is given by \cite{Guerriero2010}
\begin{align}
\Lambda_{\mathrm{B}} & \triangleq\ln\left[\frac{\int P(\bm{d}|\mathcal{H}_{1};\bm{x}_{T},\sigma_{s}^{2})\,p(\bm{x}_{T})\,p(\sigma_{s}^{2})\,d\bm{x}_{T}\,d\sigma_{s}^{2}}{P(\bm{d}|\mathcal{H}_{0})}\right]\\
 & =\ln\int\prod_{k=1}^{K}\left(\frac{P_{d,k}(\bm{x}_{T},\sigma_{s}^{2})}{P_{f,k}}\right)^{d_{k}}\left(\frac{1-P_{d,k}(\bm{x}_{T},\sigma_{s}^{2})}{1-P_{f,k}}\right)^{(1-d_{k})}\,p(\bm{x}_{T})\,p(\sigma_{s}^{2})\,d\bm{x}_{T}\,d\sigma_{s}^{2}\,.\label{eq: Totally bayesian}
\end{align}
As previously shown, the above expression can be similarly rewritten
as:
\begin{equation}
\Lambda_{\mathrm{B}}=\ln\int\exp\left(\Lambda_{{\scriptscriptstyle \mathrm{LLR}}}(\bm{x}_{T},\sigma_{s}^{2})\right)\,p(\bm{x}_{T})\,p(\sigma_{s}^{2})\,d\bm{x}_{T}\,d\sigma_{s}^{2}\,,\label{eq: Bayesian approach and LRT corresp unknown power}
\end{equation}
where $\Lambda_{{\scriptscriptstyle \mathrm{LLR}}}(\bm{x}_{T},\sigma_{s}^{2})$
has an analogous definition as in Eq. (\ref{eq: GLR - Unknown power}).

\subsubsection*{Hybrid GLRT/Bayesian approaches}

Other approaches can be obtained by mixing the two previous philosophies.
For example, assuming a prior distribution to the target position
and treating the average emitted power $\sigma_{s}^{2}$ as an unknown
and deterministic parameter, leads to the following decision statistic:
\begin{align}
\Lambda_{\mathrm{GB1}} & \triangleq\ln\left[\frac{\max_{\sigma_{s}^{2}}\int P(\bm{d}|\mathcal{H}_{1};\bm{x}_{T},\sigma_{s}^{2})\,p(\bm{x}_{T})d\bm{x}_{T}}{P(\bm{d}|\mathcal{H}_{0})}\right]\\
 & =\ln\max_{\sigma_{s}^{2}}\left\{ \int\prod_{k=1}^{K}\left(\frac{P_{d,k}(\bm{x}_{T},\sigma_{s}^{2})}{P_{f,k}}\right)^{d_{k}}\left(\frac{1-P_{d,k}(\bm{x}_{T},\sigma_{s}^{2})}{1-P_{f,k}}\right)^{(1-d_{k})}\,p(\bm{x}_{T})\,d\bm{x}_{T}\right\} \,.\label{eq: Mixed Bayesian-GLR}
\end{align}
The above statistic can be re-expressed as
\begin{equation}
\Lambda_{\mathrm{GB1}}=\max_{\sigma_{s}^{2}}\,\ln\int\exp(\Lambda_{{\scriptscriptstyle \mathrm{LLR}}}(\bm{x}_{T},\sigma_{s}^{2}))\,p(\bm{x}_{T})\,d\bm{x}_{T}\,,\label{eq: Mix Bay-GLR - LRT-dependence form}
\end{equation}
with $\Lambda_{{\scriptscriptstyle \mathrm{LLR}}}(\bm{x}_{T},\sigma_{s}^{2})$
having the usual interpretation. Alternatively, we can pursue a dual
approach, by assuming a prior distribution for $\sigma_{s}^{2}$ and
treating the target position $\bm{x}_{T}$ as unknown and deterministic.
In the latter case, the following decision statistic can be obtained:
\begin{align}
\Lambda_{\mathrm{GB2}} & \triangleq\ln\left[\frac{\max_{\bm{x}_{T}}\int P(\bm{d}|\mathcal{H}_{1};\bm{x}_{T},\sigma_{s}^{2})\,p(\sigma_{s}^{2})\,d\sigma_{s}^{2}}{P(\bm{d}|\mathcal{H}_{0})}\right]\\
 & =\ln\max_{\bm{x}_{T}}\left\{ \int\prod_{k=1}^{K}\left(\frac{P_{d,k}(\bm{x}_{T},\sigma_{s}^{2})}{P_{f,k}}\right)^{d_{k}}\left(\frac{1-P_{d,k}(\bm{x}_{T},\sigma_{s}^{2})}{1-P_{f,k}}\right)^{(1-d_{k})}\,p(\sigma_{s}^{2})\,d\sigma_{s}^{2}\right\} \,.\label{eq: Mixed GLR-Bayesian}
\end{align}
The dual statistic can be similarly rewritten as 
\begin{equation}
\Lambda_{\mathrm{GB2}}=\max_{\bm{x}_{T}}\,\ln\int\exp(\Lambda_{{\scriptscriptstyle \mathrm{LLR}}}(\bm{x}_{T},\sigma_{s}^{2}))\,p(\sigma_{s}^{2})\,d\sigma_{s}^{2}\,.\label{eq: Mix GLR-Bay - LRT-dependence form}
\end{equation}

\subsubsection*{(Hybrid) Bayesian Locally-Optimum Detection Approach}

In this case, we depart from naive Bayesian and GLRT approaches. More
specifically, our aim is to exploit the one-sided nature (when referring
to $\sigma_{s}^{2}$) of the hypothesis testing considered (cf. Eq.~(\ref{eq: Hypothesis testing -  Unknown Target Power})).
However, the problem here is complicated by the presence of the nuisance
parameter $\bm{x}_{T}$ under the hypothesis $\mathcal{H}_{1}$. To
this end, in order to get rid of the dependence on $\bm{x}_{T}$,
we consider it as an unknown random parameter and assign a prior distribution
$p(\bm{x}_{T})$. Then, we consider the averaged pdf under $\mathcal{H}_{1}$:
\begin{equation}
P(\bm{d}|\mathcal{H}_{1};\theta)=\int P(\bm{d}|\mathcal{H}_{1};\bm{x}_{T},\theta)\,p(\bm{x}_{T})\,d\bm{x}_{T}\,,\label{eq: pdf H1 Bayesian LOD}
\end{equation}
where we have used the common variable $\theta$ in the place of $\sigma_{s}^{2}$.
Once we have averaged out the dependence on $\bm{x}_{T}$, we can
apply the usual Locally-Optimum Detector (LOD), exploiting the one-sided
problem \cite{Kassam1988}. Its implicit form is given by:
\begin{equation}
\Lambda_{{\scriptscriptstyle \mathrm{BLOD}}}\triangleq\frac{\left.\frac{\partial\ln\left[P(\bm{d}|\mathcal{H}_{1};\theta)\right]}{\partial\theta}\right|_{\theta=\theta_{0}}}{\sqrt{I(\theta_{0})}},\label{eq: Bayesian LOD - general}
\end{equation}
where $I(\theta_{0})$ represents the \emph{Fisher Information} (FI)
evaluated at $\theta_{0}$, that is:
\begin{equation}
I(\theta)\triangleq\mathbb{E}\left\{ \left(\frac{\partial\ln\left[P(\bm{d}|\mathcal{H}_{1};\theta)\right]}{\partial\theta}\right)^{2}\right\} .\label{eq: Bayesian Fisher}
\end{equation}
Evaluation of the terms contained in (\ref{eq: Bayesian LOD - general})
provides the explicit form of $\Lambda_{\mathrm{{\scriptscriptstyle BLOD}}}$,
shown hereinafter (the detailed derivation is given in the Appendix):
\begin{equation}
\Lambda_{{\scriptscriptstyle \mathrm{BLOD}}}=\frac{\sum_{k=1}^{K}\frac{d_{k}-P_{f,k}}{P_{f,k}\left(1-P_{f,k}\right)}p_{w}\left(\sqrt{\frac{\gamma_{k}}{\sigma_{w,k}^{2}}}\right)\frac{\sqrt{\gamma_{k}}}{\left(\sigma_{w,k}^{2}\right)^{3/2}}\,\int g^{2}(\bm{x}_{T},\bm{x}_{k})\,p(\bm{x}_{T})d\bm{x}_{T}}{\sqrt{\sum_{k=1}^{K}\frac{1}{P_{f,k}\left(1-P_{f,k}\right)}\,p_{w}^{2}\left(\sqrt{\frac{\gamma_{k}}{\sigma_{w,k}^{2}}}\right)\frac{\gamma_{k}}{\left(\sigma_{w,k}^{2}\right)^{3}}\left(\int g^{2}(\bm{x}_{T},\bm{x}_{k})\,p(\bm{x}_{T})\,d\bm{x}_{T}\right)^{2}}}\,.\label{eq: Bayesian LOD - explicit}
\end{equation}
The so-called ``Bayesian-LOD'' (or B-LOD) statistic can be also
rewritten in a more compact form. To this end, we define the following
quantities:
\begin{align}
\nu_{k}(d_{k}) & \triangleq\frac{d_{k}-P_{f,k}}{P_{f,k}(1-P_{f,k})}\,p_{w}\left(\sqrt{\frac{\gamma_{k}}{\sigma_{w,k}^{2}}}\right)\frac{\sqrt{\gamma_{k}}}{\left(\sigma_{w,k}^{2}\right)^{3/2}}\,,\label{eq: aux def LOD 1}\\
\psi_{k} & \triangleq\frac{1}{P_{f,k}(1-P_{f,k})}\,p_{w}^{2}\left(\sqrt{\frac{\gamma_{k}}{\sigma_{w,k}^{2}}}\right)\frac{\gamma_{k}}{\left(\sigma_{w,k}^{2}\right)^{3}}\,.\label{eq: aux def LOD 2}
\end{align}
Exploiting Eqs. (\ref{eq: aux def LOD 1}) and (\ref{eq: aux def LOD 2})
into (\ref{eq: Bayesian LOD - explicit}), we obtain the equivalent
expression:
\begin{equation}
\Lambda_{{\scriptscriptstyle \mathrm{BLOD}}}=\frac{\sum_{k=1}^{K}\nu_{k}(d_{k})\,\int g^{2}(\bm{x}_{T},\bm{x}_{k})\,p(\bm{x}_{T})\,d\bm{x}_{T}}{\sqrt{\sum_{k=1}^{K}\psi_{k}\,\left(\int g^{2}(\bm{x}_{T},\bm{x}_{k})\,p(\bm{x}_{T})\,d\bm{x}_{T}\right)^{2}}}\,.\label{eq: Bayesian LOD - compact}
\end{equation}

\subsubsection*{Generalized LOD based on Davies approach}

A different approach to exploiting the one-sided nature of the problem
under investigation consists in adopting the detection approach proposed
by Davies \cite{Davies1987}. The aforementioned approach allows to
extend score-based tests to the case of nuisance parameters present
under the sole $\mathcal{H}_{1}$, as these tests require the ML estimates
of nuisances under $\mathcal{H}_{0}$ (which thus cannot be obtained).
The building rationale of Davies approach is summarized as follows. 

When $\bm{x}_{T}$ is known in (\ref{eq: Hypothesis testing -  Unknown Target Power}),
the problem reduces to a simple one-sided testing. In the latter case,
the LOD seems a reasonable decision procedure for the problem. However,
since in practice $\bm{x}_{T}$ is unknown, a \emph{family of statistics}
is rather obtained by varying $\bm{x}_{T}$. Hence, to overcome this
technical difficulty, Davies proposed the use of the \emph{maximum}
of the family of the statistics, following a ``GLRT-like'' approach.
In what follows, we will refer to the employed decision test as \emph{Generalized
LOD }(G-LOD)\emph{, }to underline the use of LOD as the inner statistic
employed in Davies approach.

The implicit form of the G-LOD is given by \cite{Davies1987}:
\begin{equation}
\Lambda_{{\scriptscriptstyle \mathrm{GLOD}}}\triangleq\max_{\bm{x}_{T}}\frac{\left.\frac{\partial\ln\left[P(\bm{d}|\mathcal{H}_{1};\bm{x}_{T},\theta)\right]}{\partial\theta}\right|_{\theta=\theta_{0}}}{\sqrt{I(\theta_{0},\bm{x}_{T})}},\label{eq: Davies LOD general}
\end{equation}
where the symbol $I(\theta,\bm{x}_{T})$ is used to denote the FI
assuming $\bm{x}_{T}$ known, that is:
\begin{equation}
I(\theta,\bm{x}_{T})\triangleq\mathbb{E}\left\{ \left(\frac{\partial\ln\left[P(\bm{d}|\mathcal{H}_{1};\bm{x}_{T},\theta)\right]}{\partial\theta}\right)^{2}\right\} .\label{eq: Fisher known x_t}
\end{equation}
The derivation of the inner term in Eq. (\ref{eq: Davies LOD general})
is provided in Appendix. The explicit form is given as:
\begin{equation}
\Lambda_{{\scriptscriptstyle \mathrm{GLOD}}}=\max_{\bm{x}_{T}}\frac{\sum_{k=1}^{K}\frac{d_{k}-P_{f,k}}{P_{f,k}(1-P_{f,k})}\,p_{w}\left(\sqrt{\frac{\gamma_{k}}{\sigma_{w,k}^{2}}}\right)\frac{\sqrt{\gamma_{k}}\,g^{2}(\bm{x}_{T},\bm{x}_{k})}{\left(\sigma_{w,k}^{2}\right)^{3/2}}}{\sqrt{\sum_{k=1}^{K}\frac{1}{P_{f,k}(1-P_{f,k})}\,p_{w}^{2}\left(\sqrt{\frac{\gamma_{k}}{\sigma_{w,k}^{2}}}\right)\frac{\gamma_{k}\,g^{4}(\bm{x}_{T},\bm{x}_{k})}{\left(\sigma_{w,k}^{2}\right)^{3}}}}\,.\label{eq: Davies LOD final}
\end{equation}
The G-LOD can be also expressed in the compact form
\begin{equation}
\Lambda_{{\scriptscriptstyle \mathrm{GLOD}}}=\max_{\bm{x}_{T}}\frac{\sum_{k=1}^{K}\nu_{k}(d_{k})\,g^{2}(\bm{x}_{T},\bm{x}_{k})}{\sqrt{\sum_{k=1}^{K}\psi_{k}\,g^{4}(\bm{x}_{T},\bm{x}_{k})}}\,,\label{eq: Davies LOD (compact form)}
\end{equation}
by exploiting the same definitions as the B-LOD in Eqs. (\ref{eq: aux def LOD 1})
and (\ref{eq: aux def LOD 2}), respectively.

\subsection{Imperfect Reporting Channels\label{sec: BSC reporting channels}}

The previous sections assumed that binary data $d_{k}$ from the WSN
could be transmitted to the FC \emph{without} any \emph{distortion}.
In this section, we consider an imperfect link scenario where the
one-bit quantized data are sent to the FC over (independent) BSCs,
in order to account for limited transmit energy and possible failures
of the sensors. We observe that the BSC model arises when separation
between sensing and communication layers is performed in the design
phase (namely a ``\emph{decode-then-fuse}'' approach \cite{Ciuonzo2012,Ciuonzo2015}). 

More specifically, we assume that the FC observes a noisy binary-valued
signal $\widehat{d}_{k}$ from $k$th sensor, that is:
\begin{equation}
\widehat{d}_{k}=\begin{cases}
d_{k} & \mathrm{with}\,\mathrm{probability\;}(1-P_{e,k})\\
1-d_{k}\quad & \mathrm{with}\,\mathrm{probability\;}P_{e,k}
\end{cases}\label{eq: BSC model}
\end{equation}
Here $P_{e,k}$ denotes the BEP on the $k$th link. Throughout this
paper we make the reasonable assumption $P_{e,k}\leq\frac{1}{2}$
and we hypothesize that $P_{e,k}$ values can be safely estimated
by the FC (that is they are \emph{known}). This is for example the
case when coherent detection or non-coherent detection with orthogonal
symbols is performed over a fading channel, as soon as the corresponding
Signal-To-Noise Ratio (SNR) can be obtained, e.g. \cite{Ciuonzo2014a,Proakis2000,Chaudhari2012}.
Then, we similarly collect the received (noisy) decisions as $\widehat{\bm{d}}\triangleq\left[\begin{array}{ccc}
\widehat{d}_{1} & \cdots & \widehat{d}_{K}\end{array}\right]^{T}$. 

It is apparent that, under hypothesis $\mathcal{H}_{1}$, the pmf
of $\widehat{\bm{d}}$ still assumes a similar (to the noise-free
reporting channels case) expression given by the product of independent
Bernoulli pmfs (since the reporting channels are assumed to act \emph{independently}),
that is:
\begin{align}
P(\widehat{\bm{d}}|\mathcal{H}_{1}) & =\prod_{k=1}^{K}P(\widehat{d}_{k}|\mathcal{H}_{1})\,=\,\prod_{k=1}^{K}(\rho_{1,k})^{\widehat{d}_{k}}\,(1-\rho_{1,k})^{(1-\widehat{d}_{k})}\,,
\end{align}
where $\rho_{1,k}\triangleq\left[\left(1-P_{e,k}\right)P_{d,k}+P_{e,k}(1-P_{d,k})\right]$.
Also, a similar expression holds for $P(\widehat{\bm{d}}|\mathcal{H}_{0})$,
when replacing $\rho_{1,k}$ with $\rho_{0,k}\triangleq\left[\left(1-P_{e,k}\right)P_{f,k}+P_{e,k}(1-P_{f,k})\right]$.
We remark that $P_{d,k}$ and $P_{f,k}$ retain the same definition
of Eq. (\ref{eq: Pd Pfa local sensors}).

\subsubsection*{Discussion: Counting Rule (CR) and Clairvoyant LLR}

First\emph{,} it is worth noticing that, the CR rule can be still
applied in the case of error-prone reporting channels, as long as
$\rho_{1,k}\geq\rho_{0,k}$. Such condition is satisfied as long as
the reasonable conditions $P_{d,k}\geq P_{f,k}$ and $P_{e,k}\leq1/2$
hold, respectively. Secondly, the (clairvoyant) LLR is given by
\begin{align}
\Lambda_{{\scriptscriptstyle \mathrm{LLR}}} & \triangleq\ln\left[\frac{P(\widehat{\bm{d}}|\mathcal{H}_{1})}{P(\widehat{\bm{d}}|\mathcal{H}_{0})}\right]=\sum_{k=1}^{K}\ln\left[\frac{P(\widehat{d}_{k}|\mathcal{H}_{1})}{P(\widehat{d}_{k}|\mathcal{H}_{0})}\right]\nonumber \\
 & =\sum_{k=1}^{K}\left\{ \widehat{d}_{k}\ln\left[\frac{\rho_{1,k}}{\rho_{0,k}}\right]+(1-\widehat{d}_{k})\ln\left[\frac{1-\rho_{1,k}}{1-\rho_{0,k}}\right]\right\} \,.\label{eq:Clairvoyant_LLR-bsc}
\end{align}
As in the case of error-free reporting channels, the clairvoyant LLR
requires knowledge of both $\bm{x}_{T}$ and $\sigma_{s}^{2}$ and
additionally of BEPs $P_{e,k}$. Also, we recall that the additionally
uncertainty arising from the BSCs should not affect the relative loss
in performance incurred by the proposed rules in comparison to the
LLR, as they will all rely on the availability of $P_{e,k}$. The
sole exception is represented by the CR, which does not rely on $P_{e,k}$s
for its implementation (it only requires $P_{e,k}\leq1/2$).

\subsubsection*{GLRT}

In the present scenario, the explicit expression of the (log-)GLR
statistic generalizes to:
\begin{align}
\Lambda_{\mathrm{G}}\triangleq & \ln\left[\frac{\max_{\sigma_{s}^{2},\bm{x}_{T}}P(\widehat{\bm{d}}|\mathcal{H}_{1};\bm{x}_{T},\sigma_{s}^{2})}{P(\widehat{\bm{d}}|\mathcal{H}_{0})}\right]\\
= & \sum_{k=1}^{K}\left\{ \widehat{d}_{k}\,\ln\left[\frac{\rho_{1,k}\left(\widehat{\bm{x}}_{T},\widehat{\sigma_{s}^{2}}\right)}{\rho_{0,k}}\right]+(1-\widehat{d}_{k})\,\ln\left[\frac{1-\rho_{1,k}\left(\widehat{\bm{x}}_{T},\widehat{\sigma_{s}^{2}}\right)}{1-\rho_{0,k}}\right]\right\} \,,\label{eq: GLRT-unknown power- BSC}
\end{align}
where $\widehat{\bm{x}}_{T}$ and $\widehat{\sigma_{s}^{2}}$ denote
the usual ML estimates of the target position and (average) emitted
reference power, assuming that $\mathcal{H}_{1}$ is true. Also, we
have adopted the notation $\rho_{1,k}(\bm{x}_{T},\sigma_{s}^{2})$
to underline the dependence on $\bm{x}_{T}$ and $\sigma_{s}^{2}$
via $P_{d,k}(\bm{x}_{T},\sigma_{s}^{2})$. Finally, we remark that
the above expression can be similarly rewritten in terms of the clairvoyant
LLR in (\ref{eq:Clairvoyant_LLR-bsc}) as Eq. (\ref{eq: GLR - Unknown power}).

\subsubsection*{Bayesian Approach\label{sub: Fully Bayesian - BSC}}

In the case of imperfect reporting channels, the explicit expression
of the (purely) Bayesian LLR generalizes to:
\begin{align}
\Lambda_{\mathrm{B}}\triangleq & \ln\left[\frac{\int P(\widehat{\bm{d}}|\mathcal{H}_{1};\bm{x}_{T},\sigma_{s}^{2})\,p(\bm{x}_{T})\,p(\sigma_{s}^{2})\,d\bm{x}_{T}\,d\sigma_{s}^{2}}{P(\widehat{\bm{d}}|\mathcal{H}_{0})}\right]\\
= & \ln\int\prod_{k=1}^{K}\left(\frac{\rho_{1,k}(\bm{x}_{T},\sigma_{s}^{2})}{\rho_{0,k}}\right)^{\widehat{d}_{k}}\times\left(\frac{1-\rho_{1,k}(\bm{x}_{T},\sigma_{s}^{2})}{1-\rho_{0,k}}\right)^{(1-\widehat{d}_{k})}p(\bm{x}_{T})\,p(\sigma_{s}^{2})\,d\bm{x}_{T}\,d\sigma_{s}^{2}\,.\label{eq: Totally bayesian BSC}
\end{align}
As previously shown, the above expression can be similarly rewritten
as in Eq. (\ref{eq: Bayesian approach and LRT corresp unknown power}),
exploiting the LLR definition provided in (\ref{eq:Clairvoyant_LLR-bsc}).

\subsubsection*{Hybrid GLRT/Bayesian approaches\label{sub:Hybrid-GLRT/Bayesian BSC}}

Hybrid GLRT/Bayesian approaches are straightforwardly extended as
follows. For example, assuming a prior for the target position $\bm{x}_{T}$
and treating $\sigma_{s}^{2}$ as deterministic provides:
\begin{align}
\Lambda_{\mathrm{GB1}} & \triangleq\ln\left[\frac{\max_{\sigma_{s}^{2}}\int P\left(\widehat{\bm{d}}|\mathcal{H}_{1};\bm{x}_{T},\sigma_{s}^{2}\right)\,p(\bm{x}_{T})d\bm{x}_{T}}{P\left(\widehat{\bm{d}}|\mathcal{H}_{0}\right)}\right]\\
 & =\ln\max_{\sigma_{s}^{2}}\int\left\{ \prod_{k=1}^{K}\left(\frac{\rho_{1,k}(\bm{x}_{T},\sigma_{s}^{2})}{\rho_{0,k}}\right)^{\widehat{d}_{k}}\left(\frac{1-\rho_{1,k}(\bm{x}_{T},\sigma_{s}^{2})}{1-\rho_{0,k}}\right)^{(1-\widehat{d}_{k})}\right\} \,p(\bm{x}_{T})\,d\bm{x}_{T}\label{eq: Mixed Bayesian-GLR - BSC}
\end{align}
The above statistic can be re-expressed in terms of the LLR similarly
as Eq. (\ref{eq: Mix Bay-GLR - LRT-dependence form}). Alternatively,
assuming a prior distribution for $\sigma_{s}^{2}$ and treating the
target position $\bm{x}_{T}$ as unknown deterministic, the complementary
hybrid statistic generalizes to:
\begin{align}
\Lambda_{\mathrm{GB2}}\triangleq & \ln\left[\frac{\max_{\bm{x}_{T}}\int P\left(\widehat{\bm{d}}|\mathcal{H}_{1};\bm{x}_{T},\sigma_{s}^{2}\right)\,p(\sigma_{s}^{2})\,d\sigma_{s}^{2}}{P\left(\widehat{\bm{d}}|\mathcal{H}_{0}\right)}\right]\\
= & \ln\max_{\bm{x}_{T}}\int\left\{ \prod_{k=1}^{K}\left(\frac{\rho_{1,k}(\bm{x}_{T},\sigma_{s}^{2})}{\rho_{0,k}}\right)^{\widehat{d}_{k}}\left(\frac{1-\rho_{1,k}(\bm{x}_{T},\sigma_{s}^{2})}{1-\rho_{0,k}}\right)^{(1-\widehat{d}_{k})}\,p(\sigma_{s}^{2})\,d\sigma_{s}^{2}\right\} \label{eq: Mixed GLR-Bayesian - BSC}
\end{align}
As usual, the above statistic can be re-expressed in terms of LLR
similarly as Eq. (\ref{eq: Mix GLR-Bay - LRT-dependence form}).

\subsubsection*{(Hybrid) Bayesian Locally-Optimum Detection Approach}

To approach the detection problem through the common LOD approach,
we first consider the averaged pdf under $\mathcal{H}_{1}$:
\begin{equation}
P\left(\widehat{\bm{d}}\,|\,\mathcal{H}_{1};\theta\right)=\int P\left(\widehat{\bm{d}}\,|\,\mathcal{H}_{1};\bm{x}_{T},\theta\right)\,p(\bm{x}_{T})\,d\bm{x}_{T}\,.
\end{equation}
The implicit form of the LOD is thus given by:
\begin{equation}
\Lambda_{{\scriptscriptstyle \mathrm{BLOD}}}\triangleq\frac{\left.\frac{\partial\ln\left[P\left(\widehat{\bm{d}}\,|\,\mathcal{H}_{1};\theta\right)\right]}{\partial\theta}\right|_{\theta=\theta_{0}}}{\sqrt{I(\theta_{0})}},\label{eq: Bayesian LOD - general (BSC)}
\end{equation}
where $I(\theta_{0})$ represents the usual FI evaluated at $\theta_{0}$,
that is:
\begin{equation}
I(\theta)\triangleq\mathbb{E}\left\{ \left(\frac{\partial\ln\left[P\left(\widehat{\bm{d}}\,|\,\mathcal{H}_{1};\theta\right)\right]}{\partial\theta}\right)^{2}\right\} .\label{eq: Bayesian Fisher - BSC}
\end{equation}
The explicit form of $\Lambda_{\mathrm{{\scriptscriptstyle BLOD}}}$
is shown as follows (the derivation is left to the reader for the
sake of brevity):
\begin{equation}
\Lambda_{{\scriptscriptstyle \mathrm{BLOD}}}=\frac{\sum_{k=1}^{K}\frac{\widehat{d}_{k}-\rho_{0,k}}{\rho_{0,k}\left(1-\rho_{0,k}\right)}\,(1-2P_{e,k})\,p_{w}\left(\sqrt{\frac{\gamma_{k}}{\sigma_{w,k}^{2}}}\right)\frac{\sqrt{\gamma_{k}}}{\left(\sigma_{w,k}^{2}\right)^{3/2}}\left(\int g^{2}(\bm{x}_{T},\bm{x}_{k})\,p(\bm{x}_{T})\,d\bm{x}_{T}\right)}{\sqrt{\sum_{k=1}^{K}\frac{1}{P_{f,k}\left(1-P_{f,k}\right)}\,p_{w}^{2}\left(\sqrt{\frac{\gamma_{k}}{\sigma_{w,k}^{2}}}\right)\frac{\gamma_{k}}{\left(\sigma_{w,k}^{2}\right)^{3}}\left(\int g^{2}(\bm{x}_{T},\bm{x}_{k})\,p(\bm{x}_{T})\,d\bm{x}_{T}\right)^{2}}}\,.\label{eq: Bayesian LOD - explicit- BSC}
\end{equation}
Similarly, by exploiting the following generalized definitions
\begin{align}
\widehat{\nu}_{k}(\widehat{d}_{k}) & \triangleq\frac{\widehat{d}_{k}-\rho_{0,k}}{\rho_{0,k}\left(1-\rho_{0,k}\right)}\,(1-2P_{e,k})\,p_{w}\left(\sqrt{\frac{\gamma_{k}}{\sigma_{w,k}^{2}}}\right)\frac{\sqrt{\gamma_{k}}}{\left(\sigma_{w,k}^{2}\right)^{3/2}}\,,\label{eq: aux def LOD 1- BSC}\\
\widehat{\psi}_{k} & \triangleq\frac{1}{\rho_{0,k}\left(1-\rho_{0,k}\right)}\,(1-2P_{e,k})^{2}\,p_{w}^{2}\left(\sqrt{\frac{\gamma_{k}}{\sigma_{w,k}^{2}}}\right)\frac{\gamma_{k}}{\left(\sigma_{w,k}^{2}\right)^{3}}\,,\label{eq: aux def LOD 2- BSC}
\end{align}
the B-LOD can be also expressed in a similar compact form:
\begin{equation}
\Lambda_{{\scriptscriptstyle \mathrm{BLOD}}}=\frac{\sum_{k=1}^{K}\widehat{\nu}_{k}(\widehat{d}_{k})\,\int g^{2}(\bm{x}_{T},\bm{x}_{k})\,p(\bm{x}_{T})\,d\bm{x}_{T}}{\sqrt{\sum_{k=1}^{K}\widehat{\psi}_{k}\,\left(\int g^{2}(\bm{x}_{T},\bm{x}_{k})\,p(\bm{x}_{T})\,d\bm{x}_{T}\right)^{2}}}\,.\label{eq: Bayesian LOD - compact - BSC}
\end{equation}

\subsubsection*{Generalized LOD based on Davies approach}

The implicit form of LOD based on Davies approach is given by \cite{Davies1987}:
\begin{equation}
\Lambda_{{\scriptscriptstyle \mathrm{GLOD}}}\triangleq\max_{\bm{x}_{T}}\frac{\left.\frac{\partial\ln\left[P\left(\widehat{\bm{d}}\,|\mathcal{H}_{1};\bm{x}_{T},\theta\right)\right]}{\partial\theta}\right|_{\theta=\theta_{0}}}{\sqrt{I(\bm{x}_{T},\theta_{0})}},\label{eq: Davies LOD general-1}
\end{equation}
where the symbol $I(\bm{x}_{T},\theta)$ is used to denote the FI
assuming $\bm{x}_{T}$ known, that is:
\begin{equation}
I(\bm{x}_{T},\theta)\triangleq\mathbb{E}\left\{ \left(\frac{\partial\ln\left[P(\widehat{\bm{d}}|\mathcal{H}_{1};\bm{x}_{T},\theta)\right]}{\partial\theta}\right)^{2}\right\} .\label{eq: Fisher known x_t - BSC}
\end{equation}
The derivation of the inner term in Eq. (\ref{eq: Davies LOD general-1})
is left to the reader for sake of brevity. The explicit form is given
as:
\begin{equation}
\Lambda_{{\scriptscriptstyle \mathrm{GLOD}}}=\max_{\bm{x}_{T}}\frac{\sum_{k=1}^{K}\frac{\widehat{d}_{k}-\rho_{0,k}}{\rho_{0,k}\left(1-\rho_{0,k}\right)}\,(1-2P_{e,k})\,p_{w}\left(\sqrt{\frac{\gamma_{k}}{\sigma_{w,k}^{2}}}\right)\frac{\sqrt{\gamma_{k}}\,g^{2}(\bm{x}_{T},\bm{x}_{k})}{\left(\sigma_{w,k}^{2}\right)^{3/2}}}{\sqrt{\sum_{k=1}^{K}\frac{(1-2P_{e,k})^{2}}{\rho_{0,k}\left(1-\rho_{0,k}\right)}\,p_{w}^{2}\left(\sqrt{\frac{\gamma_{k}}{\sigma_{w,k}^{2}}}\right)\frac{\gamma_{k}\,g^{4}(\bm{x}_{T},\bm{x}_{k})}{\left(\sigma_{w,k}^{2}\right)^{3}}}}\,.\label{eq: Davies LOD final- BSC}
\end{equation}
Similarly, G-LOD can be also expressed in the compact form:
\begin{equation}
\Lambda_{{\scriptscriptstyle \mathrm{GLOD}}}=\max_{\bm{x}_{T}}\frac{\sum_{k=1}^{K}\widehat{\nu}_{k}(\widehat{d}_{k})\,g^{2}(\bm{x}_{T},\bm{x}_{k})}{\sqrt{\sum_{k=1}^{K}\widehat{\psi}_{k}\,\,g^{4}(\bm{x}_{T},\bm{x}_{k})}}\,,\label{eq: Davies LOD (compact form) - BSC}
\end{equation}
by exploiting the same definitions as the B-LOD in Eqs. (\ref{eq: aux def LOD 1- BSC})
and (\ref{eq: aux def LOD 2- BSC}), respectively.

\subsection{Summary of the Considered Rules and their Practical Implementation\label{sub: Complexity comparison}}

In this section, we provide a summarizing comparison of the considered
rules, focusing on the computational complexity (a performance comparison
is then provided in Sec. \ref{sec: Simulation results}). To this
end, in Tab. \ref{tab: Detectors comparison} we report the explicit
form of the considered fusion rules, as well as the corresponding
complexity required for their implementation. 

First of all, we observe that CR (cf. Eq. (\ref{eq:Counting_rule}))
and B-LOD (cf. Eq. (\ref{eq: Bayesian LOD - compact - BSC})) require
the lowest complexity (that is $\mathcal{O}(K)$), as only a sum of
$K$ terms needs to be evaluated (indeed the integrations of B-LOD
in (\ref{eq: Bayesian LOD - compact - BSC}) can be performed off-line).
Secondly, all the remaining rules require optimizations (GLRT and
G-LOD), integrations (Bayesian approach) or both of them (viz. hybrid
approaches). In this case, the complexity evaluation in Tab. \ref{tab: Detectors comparison}
subsumes that a \emph{grid search or integration} is performed, similarly
as in \cite{Niu2006b,Guerriero2010,Shoari2012}. Additionally, when
dealing with prior pdfs, we employ non-informative priors with the
intent of underlining useful analogies among proposed rules. Nonetheless,
grid implementation (and corresponding complexity evaluation) still
applies to the case of informative priors. 

More specifically, after assuming that $\bm{x}_{T}$ and $\sigma_{s}^{2}$
belong to limited sets $\mathrm{S}_{\bm{x}_{T}}\subset\mathbb{R}^{d}$
and $S_{\mathrm{\sigma_{s}^{2}}}\subset\mathbb{R}^{+}$, respectively,
the space ($\bm{x}_{T},\sigma_{s}^{2})$ is then discretized into:
\begin{itemize}
\item $N_{\bm{x}_{T}}$ position bins in the $d$-dimensional space, each
one associated to a center bin position, say $\bm{x}_{T}[i]$, $i\in\{1,\ldots N_{\bm{x}_{T}}\}$;
\item $N_{\sigma_{s}^{2}}$ variance (power) bins, each one to associated
to a center bin variance, say $\sigma_{s}^{2}[j]$, $j\in\{1,\ldots N_{\sigma_{s}^{2}}\}$.
\end{itemize}
\emph{Grid implementation of GLRT:} Starting from the alternative
form of Eq. (\ref{eq: GLR - Unknown power}), we approximate the GLRT
via the following grid search:
\begin{equation}
\Lambda_{\mathrm{G}}\approx\max_{i=1,\ldots N_{\bm{x}_{T}}}\,\max_{j=1,\ldots,N_{\sigma_{s}^{2}}}\,\Lambda_{{\scriptscriptstyle \mathrm{LLR}}}(\bm{x}_{T}[i]\,,\,\sigma_{s}^{2}[j])\,;\label{eq: GLRT grid discretization}
\end{equation}
where we recall that $\Lambda_{{\scriptscriptstyle \mathrm{LLR}}}(\bm{x}_{T}[i]\,,\,\sigma_{s}^{2}[j])$\emph{
}represents the expression of the clairvoyant LLR statistic obtained
by evaluating the $P_{d,k}$s by replacing $\bm{x}_{T}$ and $\sigma_{s}^{2}$
with $\bm{x}_{T}[i]$ and $\sigma_{s}^{2}[j]$, respectively, into
(\ref{eq: Pd Pfa local sensors}).

\emph{Grid implementation of Bayesian approach: }First, we approximate
the double integral in (\ref{eq: Bayesian approach and LRT corresp unknown power})
through the Riemann sums as follows:
\begin{align}
\Lambda_{\mathrm{B}} & \approx\ln\left[\sum_{i=1}^{N_{\bm{x}_{T}}}\,\sum_{j=1}^{N_{\sigma_{s}^{2}}}\exp\left\{ \Lambda_{{\scriptscriptstyle \mathrm{LLR}}}(\bm{x}_{T}[i],\sigma_{s}^{2}[j])+\ln\,r_{i}+\ln\,\bar{r}_{j}\right\} \right]\,,\label{eq: Riemann Bayesian - unknown power}
\end{align}
where $r_{i}$ and $\bar{r}_{j}$ are the the mass probabilities associated
to bins $i$ and $j$ of $\bm{x}_{T}$ and $\sigma_{s}^{2}$, through
$p(\bm{x}_{T})$ and $p(\sigma_{s}^{2})$, respectively. In other
words, $r_{i}\triangleq\Pr\{\bm{x}_{T}\in\mathcal{I}(\bm{x}_{T}[i])\}$
and $\bar{r}_{j}\triangleq\Pr\{\sigma_{s}^{2}\in\mathcal{I}(\sigma_{s}^{2}[j])\}$,
where $\mathcal{I}(\bm{x}_{T}[i])$ and $\mathcal{I}(\sigma_{s}^{2}[j])$
denote the extent of $i$th and $j$th bins of the grid employed.
This approximation admits a more intuitive form when the prior pdfs
are assumed non-informative (viz. uniform). Indeed, in the latter
case, the above approximation specializes into:
\begin{align}
\Lambda_{\mathrm{B}} & \approx r\,\bar{r}\,\ln\left[\sum_{i=1}^{N_{\bm{x}_{T}}}\,\sum_{j=1}^{N_{\sigma_{s}^{2}}}\,\exp(\Lambda_{{\scriptscriptstyle \mathrm{LLR}}}(\bm{x}_{T}[i],\sigma_{s}^{2}[j]))\right]\,,\\
 & \propto\ln\left[\sum_{i=1}^{N_{\bm{x}_{T}}}\,\sum_{j=1}^{N_{\sigma_{s}^{2}}}\,\exp(\Lambda_{{\scriptscriptstyle \mathrm{LLR}}}(\bm{x}_{T}[i],\sigma_{s}^{2}[j]))\right]\,.\label{eq: Bayesian - Riemann uniform pdf}
\end{align}
The right-hand side is in the form of the well-known log-sum-exp combination,
which can be also interpreted as a ``soft-max'' function. Therefore
it is apparent that GLR approximation in (\ref{eq: GLRT grid discretization})
shows a clear connection with the Bayesian approach in (\ref{eq: Bayesian - Riemann uniform pdf}),
as also observed in \cite{Guerriero2010} for the case of random sensor
deployment.

\emph{Grid implementation of hybrid approaches:} Remarkably, the hybrid
fusion rules of Sec. \ref{sub:Hybrid-GLRT/Bayesian BSC} admit similar
approximations as the pure GLR and Bayesian decision statistic. Indeed,
$\Lambda_{\mathrm{GB1}}$ in (\ref{eq: Mix Bay-GLR - LRT-dependence form})
is approximated (assuming a uniform pdf for $p(\bm{x}_{T})$) as
\begin{equation}
\Lambda_{\mathrm{GB1}}\approx\max_{j=1,\ldots,N_{\sigma_{s}^{2}}}\,\ln\sum_{i=1}^{N_{\bm{x}_{T}}}\exp(\Lambda_{{\scriptscriptstyle \mathrm{LLR}}}(\bm{x}_{T}[i],\sigma_{s}^{2}[j]))\,,\label{eq: GB1 Riemann sums}
\end{equation}
while $\Lambda_{\mathrm{GB2}}$ in (\ref{eq: Mix GLR-Bay - LRT-dependence form}) is
approximated (assuming a uniform pdf for $p(\sigma_{s}^{2})$) as:
\begin{equation}
\Lambda_{\mathrm{GB2}}\approx\max_{i=1,\ldots N_{\bm{x}_{T}}}\,\ln\sum_{j=1}^{N_{\sigma_{s}^{2}}}\exp(\Lambda_{{\scriptscriptstyle \mathrm{LLR}}}(\bm{x}_{T}[i],\sigma_{s}^{2}[j]))\,.\label{eq: GB2 Riemann sums}
\end{equation}
The above expressions underline the soft-max approach with respect
to one variable and a max approach with respect to the other. Clearly,
the computational complexity of all these methods is based on the
evaluation of the statistic at the grid points, thus implying $\mathcal{O}\left(K\cdot N_{\bm{x}_{T}}\cdot N_{\sigma_{s}^{2}}\right)$. 

\emph{Grid implementation of G-LOD:} Finally, the G-LOD can be approximated
in a similar way by discretizing \emph{only} the search space of $\bm{x}_{T}$
as:
\begin{equation}
\Lambda_{{\scriptscriptstyle \mathrm{GLOD}}}\approx\max_{i=1,\ldots N_{\bm{x}_{T}}}\frac{\sum_{k=1}^{K}\widehat{\nu}_{k}(\widehat{d}_{k})\,g^{2}(\bm{x}_{T}[i],\bm{x}_{k})}{\sqrt{\sum_{k=1}^{K}\widehat{\psi}_{k}\,\,g^{4}(\bm{x}_{T}[i],\bm{x}_{k})}}\,.\label{eq: Davies LOD Riemann Sums}
\end{equation}
Therefore, its complexity is given by $\mathcal{O}\left(K\cdot N_{\bm{x}_{T}}\right)$
and provides a \emph{dramatic} \emph{reduction} in \emph{complexity}
with respect to other rules based on grid implementation.

\begin{table*}
\begin{centering}
\begin{tabular}{c||c|c}
\hline 
\textbf{Fusion Rule} & \textbf{Explicit Expression} & \textbf{Computational Complexity}\tabularnewline
\hline 
\hline 
\noalign{\vskip0.3cm}
GLR & $\max_{\bm{x}_{T},\sigma_{s}^{2}}\,\Lambda_{{\scriptscriptstyle \mathrm{LLR}}}(\bm{x}_{T},\sigma_{s}^{2})$ & $\mathcal{O}\left(K\cdot N_{\bm{x}_{T}}\cdot N_{\sigma_{s}^{2}}\right)$
(Grid)\tabularnewline[0.3cm]
\hline 
\hline 
\noalign{\vskip0.3cm}
Bayesian & $\ln\int\exp(\Lambda_{{\scriptscriptstyle \mathrm{LLR}}}(\bm{x}_{T},\sigma_{s}^{2}))\,p(\bm{x}_{T})\,p(\sigma_{s}^{2})d\bm{x}_{T}\,d\sigma_{s}^{2}$ & $\mathcal{O}\left(K\cdot N_{\bm{x}_{T}}\cdot N_{\sigma_{s}^{2}}\right)$
(Grid)\tabularnewline[0.3cm]
\hline 
\noalign{\vskip0.3cm}
Hybrid Approach 1 & $\max_{\sigma_{s}^{2}}\,\ln\int\exp(\Lambda_{{\scriptscriptstyle \mathrm{LLR}}}(\bm{x}_{T},\sigma_{s}^{2}))\,p(\bm{x}_{T})\,d\bm{x}_{T}$ & $\mathcal{O}\left(K\cdot N_{\bm{x}_{T}}\cdot N_{\sigma_{s}^{2}}\right)$
(Grid)\tabularnewline[0.3cm]
\hline 
\noalign{\vskip0.3cm}
Hybrid Approach 2 & $\max_{\bm{x}_{T}}\,\ln\int\exp(\Lambda_{{\scriptscriptstyle \mathrm{LLR}}}(\bm{x}_{T},\sigma_{s}^{2}))\,p(\sigma_{s}^{2})\,d\sigma_{s}^{2}$ & $\mathcal{O}\left(K\cdot N_{\bm{x}_{T}}\cdot N_{\sigma_{s}^{2}}\right)$
(Grid)\tabularnewline[0.3cm]
\hline 
\noalign{\vskip0.3cm}
Bayesian LOD & $\frac{\sum_{k=1}^{K}\widehat{\nu}_{k}(\widehat{d}_{k})\,\int g^{2}(\bm{x}_{T},\bm{x}_{k})\,p(\bm{x}_{T})d\bm{x}_{T}}{\sqrt{\sum_{k=1}^{K}\widehat{\psi}_{k}\left(\int g^{2}(\bm{x}_{T},\bm{x}_{k})\,p(\bm{x}_{T})\,d\bm{x}_{T}\right)^{2}}}$ & $\mathcal{O}(K)$\tabularnewline[0.3cm]
\hline 
\noalign{\vskip0.3cm}
Counting Rule & $\sum_{k=1}^{K}\widehat{d}_{k}$ & $\mathcal{O}(K)$\tabularnewline[0.3cm]
\hline 
\noalign{\vskip0.3cm}
Generalized LOD & $\max_{\bm{x}_{T}}\frac{\sum_{k=1}^{K}\widehat{\nu}_{k}(\widehat{d}_{k})\,\,g^{2}(\bm{x}_{T},\bm{x}_{k})}{\sqrt{\sum_{k=1}^{K}\widehat{\psi}_{k}\,g^{4}(\bm{x}_{T},\bm{x}_{k})}}$ & $\mathcal{O}\left(K\cdot N_{\bm{x}_{T}}\right)$ (Grid)\tabularnewline[0.3cm]
\hline 
\end{tabular}
\par\end{centering}

\caption{Comparison of decision statistics; $\Lambda_{{\scriptscriptstyle \mathrm{LLR}}}(\bm{x}_{T})$
and $\Lambda_{{\scriptscriptstyle \mathrm{LLR}}}(\bm{x}_{T},\sigma_{s}^{2})$
are defined through Eq. (\ref{eq:Clairvoyant_LR}).\label{tab: Detectors comparison}}
\end{table*}

\section{Simulation Results\label{sec: Simulation results}}

\begin{figure}
\centering{}\includegraphics[width=0.6\paperwidth]{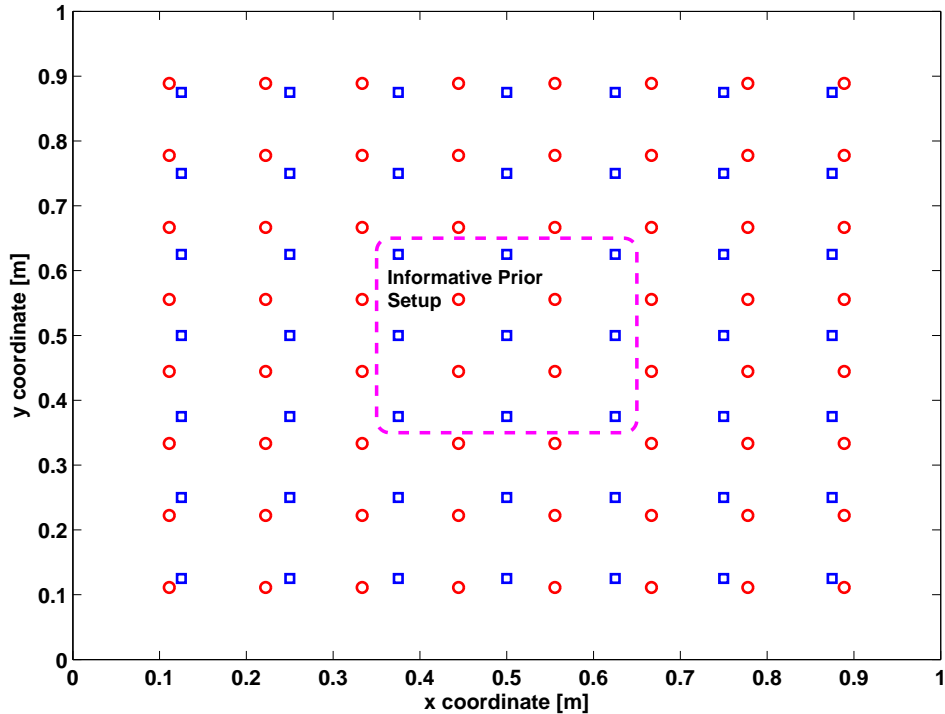}\caption{Regular deployment of WSN in the case of $K=49$ (blue ``$\square$''
markers) and $K=64$ (red ``$\circ$'' markers) sensors. The area
delimited by magenta dashed line refers to informative prior setup,
i.e. $\mathrm{S}_{\bm{x}_{T}}\triangleq[0.35,0.65]\times[0.35,0.65]$.\label{fig: Regular WSN deployment}}
\end{figure}

In this section we compare the performance of the considered rules
through numerical results. To this end, we consider a 2-D scenario
($\bm{x}_{T}\in\mathbb{R}^{2}$) where a WSN is employed to detect
the presence of a target within the region $[0,1]\times[0,1]$, which
represents the considered surveillance area. The sensors are arranged
according to a regular square grid covering the surveillance area,
as shown in Fig. \ref{fig: Regular WSN deployment}, where two cases
concerning $K=49$ and $K=64$ sensors are illustrated.

With reference to the sensing model, for simplicity we assume the
same measurement variance for all the sensors, i.e. $\sigma_{w,k}^{2}=\sigma_{w}^{2}$.
Also, without loss of generality, we set $\sigma_{w}^{2}=1$. Differently,
with reference to the AAF, we will both consider both the cases of ($i$)
power-law and ($ii$) exponential AAFs, with parameter values: $\eta=0.2$
(viz. approximate target extent); $\alpha=4$ (power-law AAF decay
exponent). Finally, we define the local sensing SNR as $\mathrm{SNR}\triangleq10\,\log_{10}\frac{\sigma_{s}^{2}}{\sigma_{w}^{2}}$.
The local false-alarm rate for every sensor is set to $P_{f,k}=0.05$
(the corresponding decision threshold $\gamma_{k}$ is obtained by
inverting relationship in Eq. (\ref{eq: Pd Pfa local sensors})).
When not otherwise specified, we assume ideal reporting channels,
i.e. $P_{e,k}=0$, $k\in\mathcal{K}$.

With reference to grid-based approaches (cf. Sec. \ref{sub: Complexity comparison}),
those employed for $\bm{x}_{T}$ and $\sigma_{s}^{2}$ are the following:
\begin{itemize}
\item Target position $\bm{x}_{T}$: the search (resp. integration) space
corresponds to the surveillance area, i.e. $\mathrm{S}_{\bm{x}_{T}}=[0,1]\times[0,1]$.
The x-and y-coordinates grid spacings are given by $(1/N_{x})$, where
$N_{x}=100$ is chosen here;
\item Target average emitted power $\sigma_{s}^{2}$: the search (resp.
integration) space is chosen as 
\begin{equation}
S_{\mathrm{\sigma_{s}^{2}}}=\left[(1-\rho_{s})\cdot\sigma_{s,\mathrm{true}}^{2}\:,\:(1+\rho_{s})\cdot\sigma_{s,\mathrm{true}}^{2}\right]\,,
\end{equation}
where $\sigma_{s,\mathrm{true}}^{2}$ denotes the emitted power true
value and $\rho_{s}=\frac{1}{10}$, which provides a relative $20\%$
uncertainty with respect to $\sigma_{s,\mathrm{true}}^{2}$. The grid
spacing is given by $\frac{2\rho_{s}\sigma_{s,\mathrm{true}}^{2}}{N_{\sigma}}$,
where $N_{\sigma}=10$ is chosen here;
\end{itemize}
In what follows we compare the considered rules through their corresponding
ROCs based on Monte Carlo simulations, obtained with $10^{5}$ runs.
The ROC performance reported refer to a scenario where $\bm{x}_{T}$
is uniformly randomly generated at each run within the surveillance
area $\mathrm{S}_{\bm{x}_{T}}$.

\begin{figure}
\centering{}\subfloat[Power-law AAF.]{\centering{}\includegraphics[width=0.42\paperwidth]{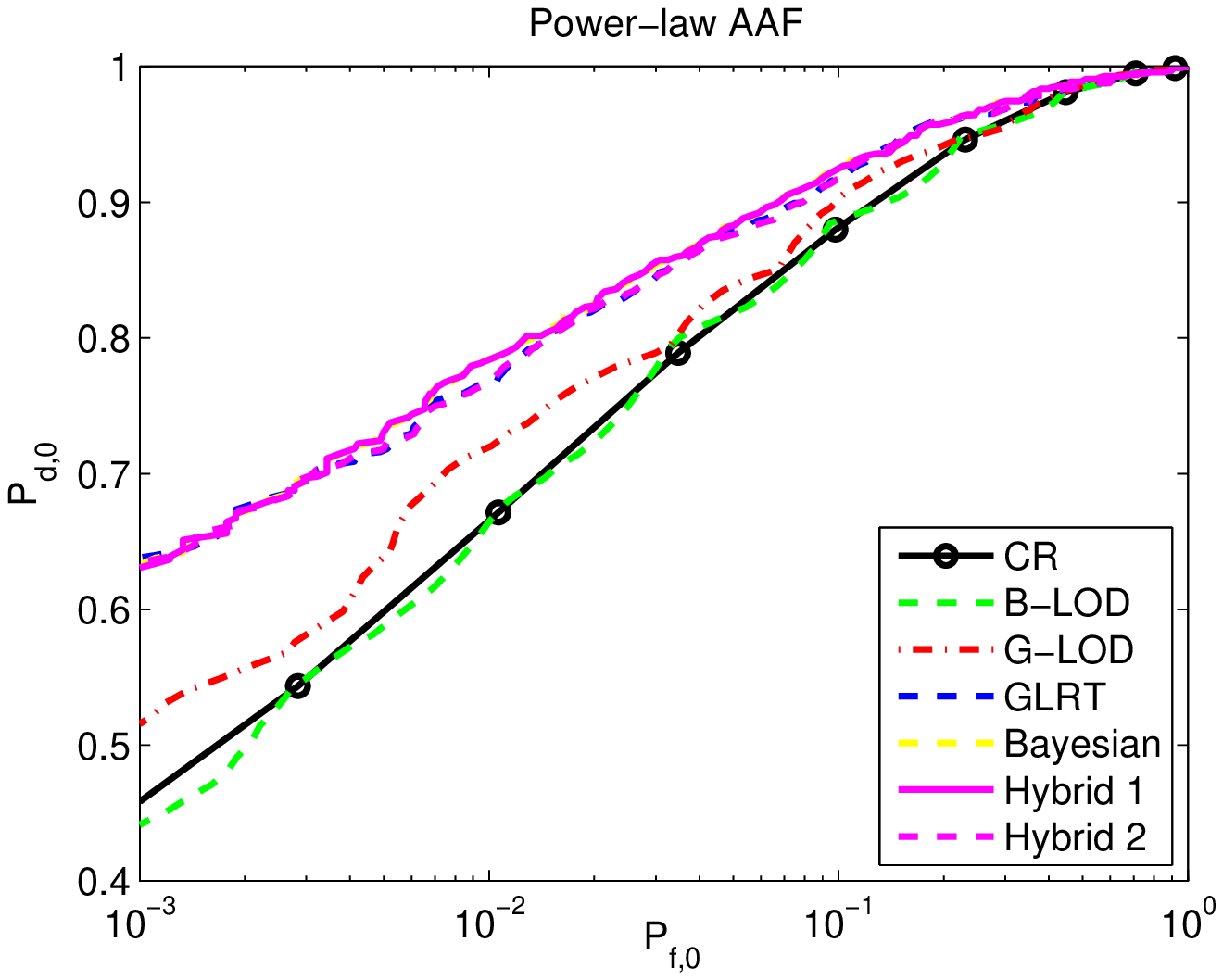}}\subfloat[Exponential AAF.]{\centering{}\includegraphics[width=0.42\paperwidth]{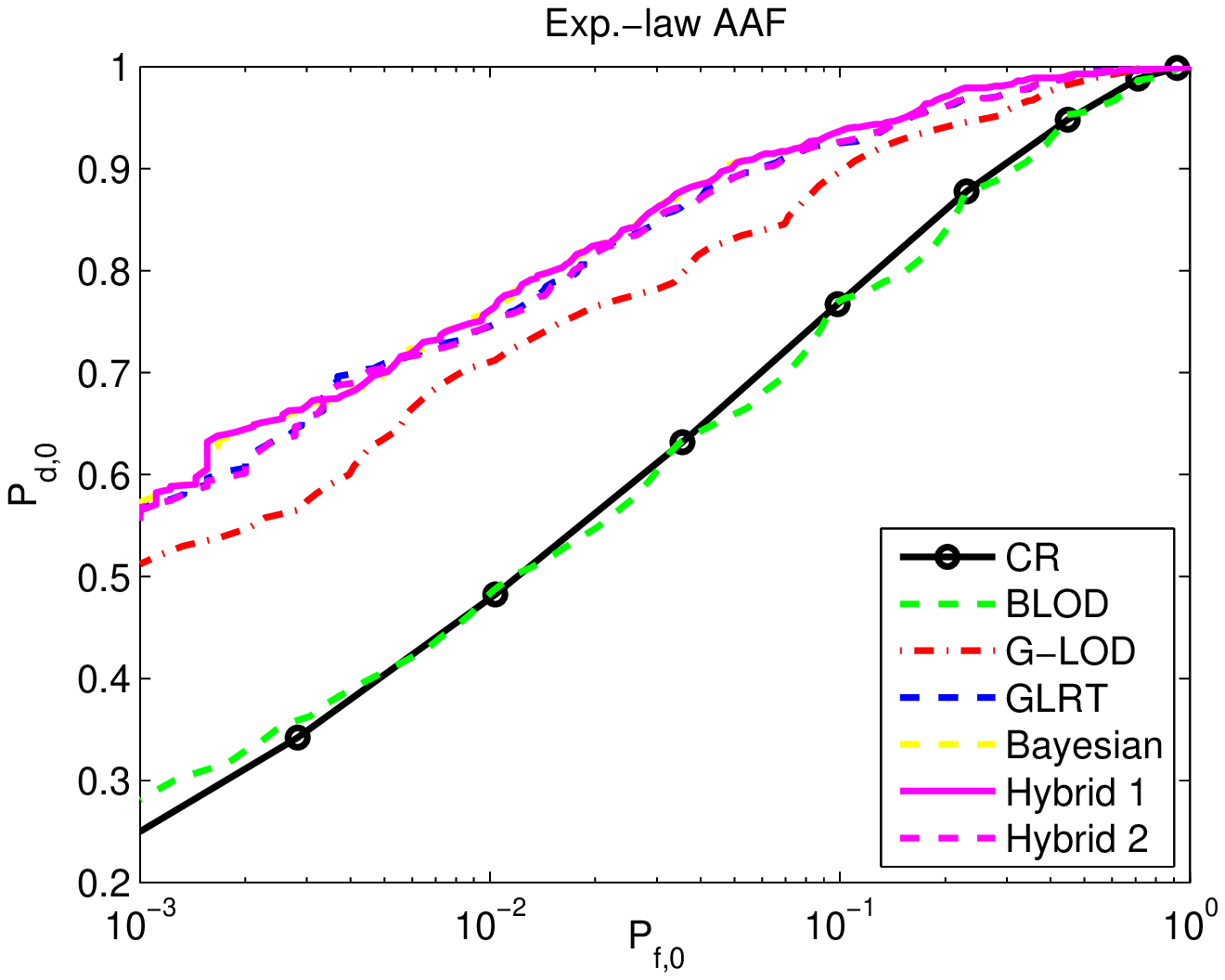}}\caption{$P_{d,0}$ vs. $P_{f,0}$ for all the presented rules; WSN with $K=49$
sensors, $\mathrm{SNR}=10\,\mathrm{dB}$, $P_{e,k}=0$ (ideal reporting
channels).\label{fig: ROC ideal BEP K =00003D 49}}
\end{figure}

First, in Fig. \ref{fig: ROC ideal BEP K =00003D 49} we report the
ROCs for both the cases of power-law AAF (subfigure ($a$)) and exponential
AAF (subfigure ($b$)) in a WSN with $K=49$ sensors arranged as in
Fig. \ref{fig: Regular WSN deployment} (blue ``$\square$'' markers).
First of all, it is apparent that B-LOD and CR achieve almost the
same performance in this scenario. Therefore, the prior information
on $\bm{x}_{T}$ is too vague and does not provide itself a relevant
gain w.r.t. ``blind assumption'' of CR, which also arises from different
founding rationales (cf. Sec. \ref{sub: Decision Fusion}). Differently,
all the other rules achieve a significant performance improvement
over CR. Moreover, purely Bayesian and GLRT approaches, as well as
the hybrid ones, roughly achieve the same performance under both power-law
and exponential AAFs. Interestingly, G-LOD achieves a worth performance
gain w.r.t. CR, especially in the case of an exponential AAF. This
is motivated by a faster signal decay (viz. a more ``sensitive''
spatial signature), which is effectively exploited by the maximization
required for G-LOD implementation (see Eq. (\ref{eq: Davies LOD (compact form) - BSC})).
Also, from inspection of the figures, G-LOD suffers from a slight
performance loss when compared to remaining grid-based approaches.
However, such loss is balanced by a significant lower complexity required,
thus confirming its attractiveness.

\begin{figure}
\centering{}\subfloat[Power-law AAF.]{\centering{}\includegraphics[width=0.42\paperwidth]{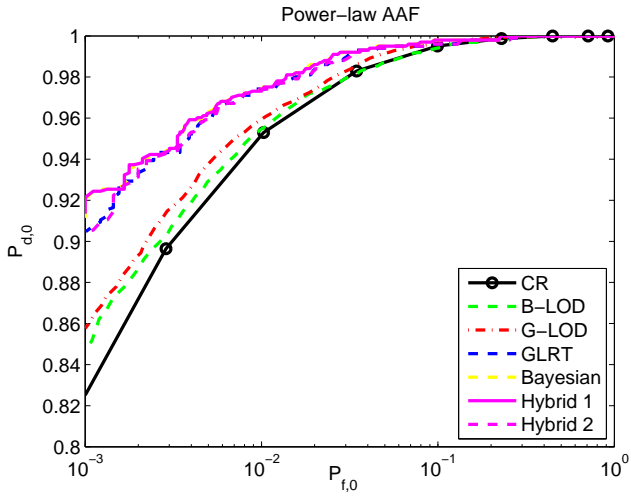}}\subfloat[Exponential AAF.]{\centering{}\includegraphics[width=0.42\paperwidth]{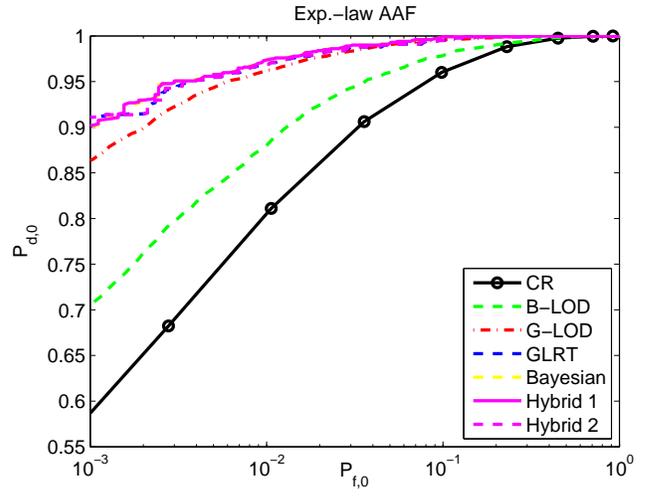}}\caption{$P_{d,0}$ vs. $P_{f,0}$ for all the presented rules (informative
prior setup); WSN with $K=49$ sensors, $\mathrm{SNR}=10\,\mathrm{dB}$,
$P_{e,k}=0$ (ideal reporting channels).\label{fig: ROC ideal BEP K =00003D 49 (informative setup)}}
\end{figure}

Differently, in Fig. \ref{fig: ROC ideal BEP K =00003D 49 (informative setup)}
we report similar ROCs for the case of a more informative prior availability
on $\bm{x}_{T}$. More specifically, we assume that $\bm{x}_{T}\in\mathrm{S}_{\bm{x}_{T}}\triangleq[0.35,0.65]\times[0.35,0.65]$
(see Fig. \ref{fig: Regular WSN deployment}), i.e. the target can
be located (when present) within a smaller square than the WSN deployment
region (see Fig. \ref{fig: Regular WSN deployment}). By looking at
the figure, similar considerations can be also drawn in this setup,
except for the ROC performance achieved by B-LOD. Indeed, in the latter
case, the exploitation of the more informative prior pdf $p(\bm{x}_{T})$
available (i.e. a uniform one on a smaller area) overcomes the blind
nature behind CR derivation. Therefore, when accurate information
on target potential position is available, B-LOD represents an interesting
alternative rule, since its complexity grows only \emph{linearly}
with the number of sensors $K$ (cf. Tab. \ref{tab: Detectors comparison}).

\begin{figure}
\centering{}\subfloat[Power-law AAF.]{\centering{}\includegraphics[width=0.42\paperwidth]{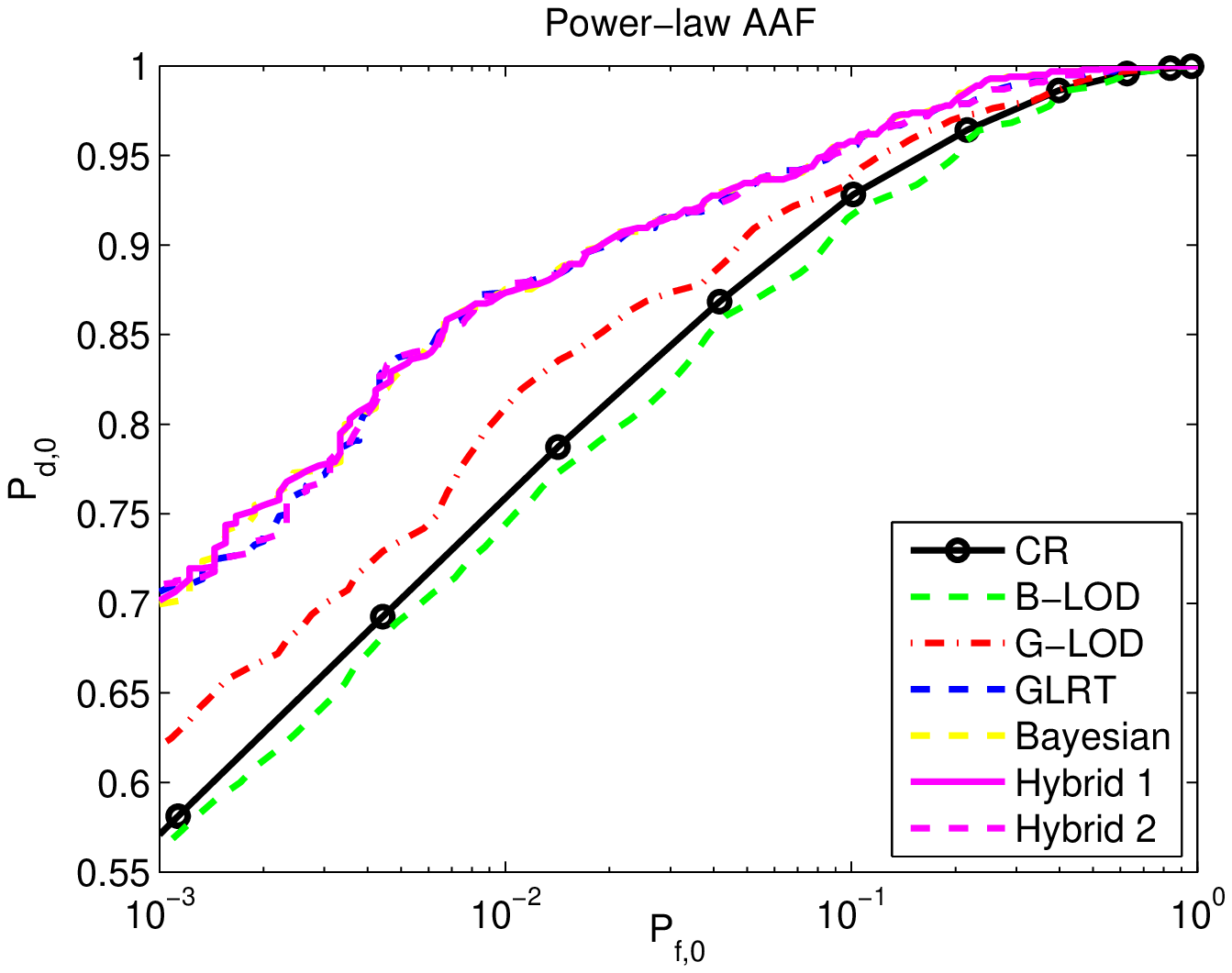}}\subfloat[Exponential AAF.]{\centering{}\includegraphics[width=0.42\paperwidth]{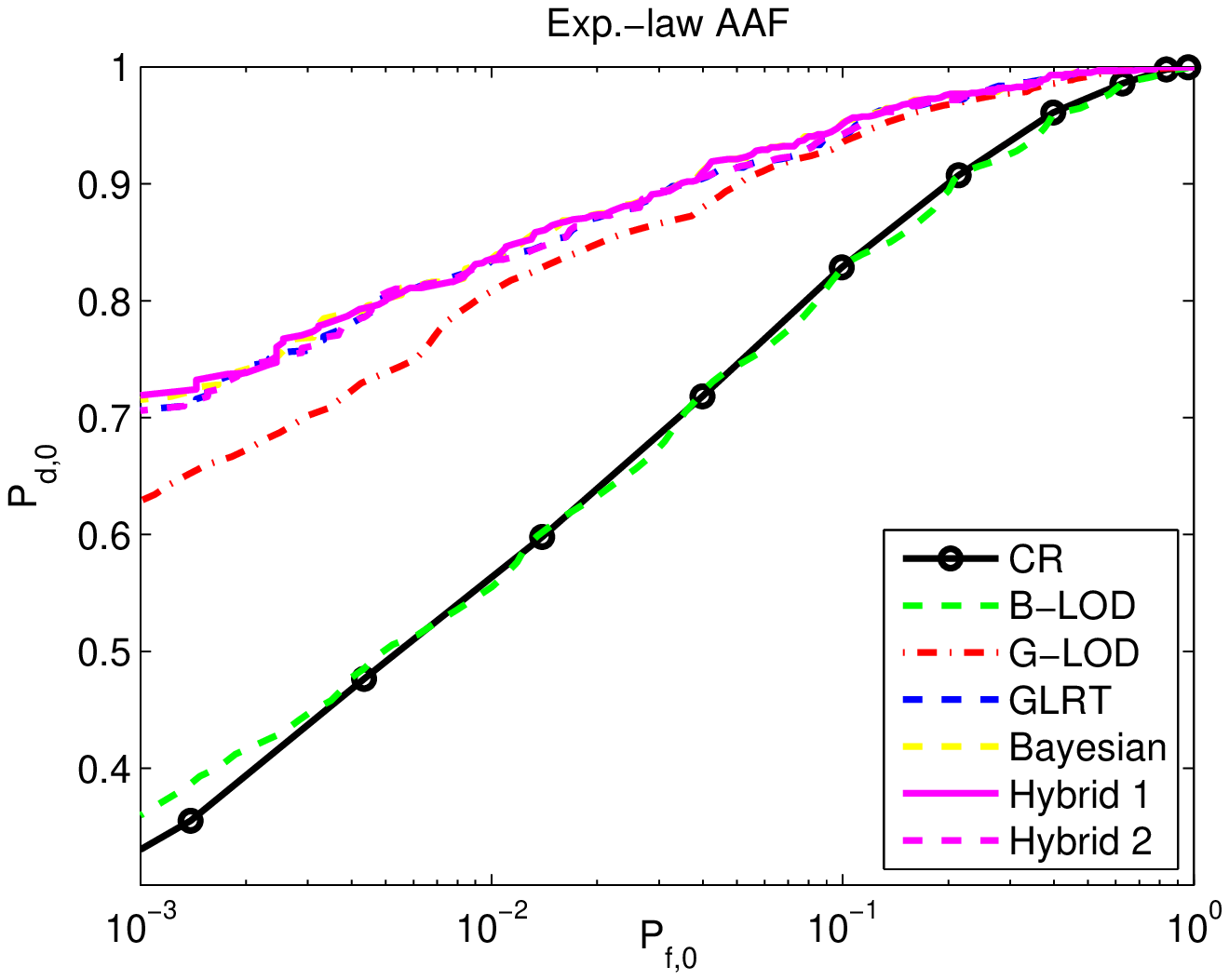}}\caption{$P_{d,0}$ vs. $P_{f,0}$ for all the presented rules; WSN with $K=64$
sensors, $\mathrm{SNR}=10\,\mathrm{dB}$, $P_{e,k}=0$ (ideal reporting
channels).\label{fig: ROC ideal BEP K =00003D 64}}
\end{figure}

\begin{figure}
\centering{}\subfloat[Power-law AAF.]{\centering{}\includegraphics[width=0.42\paperwidth]{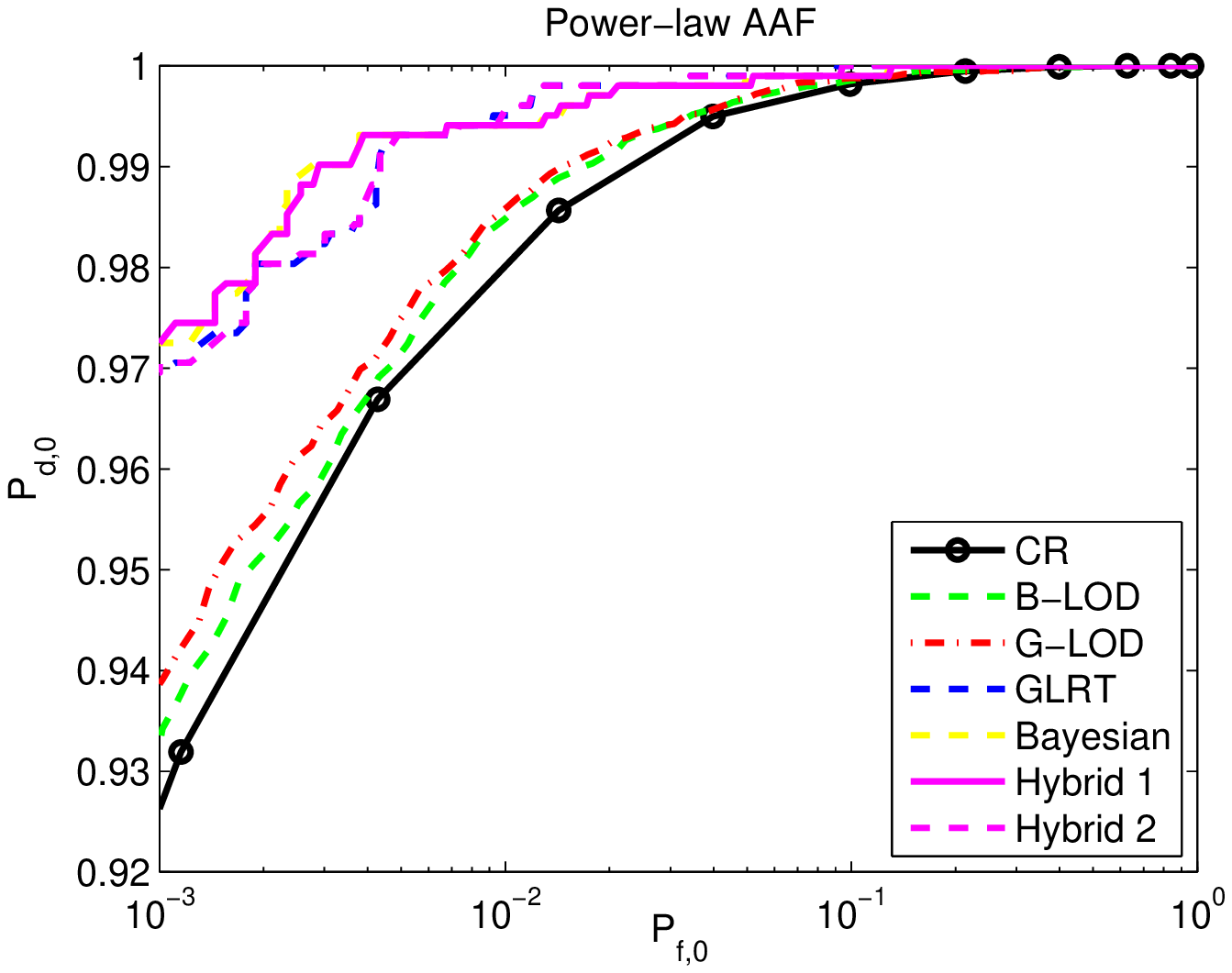}}\subfloat[Exponential AAF.]{\centering{}\includegraphics[width=0.42\paperwidth]{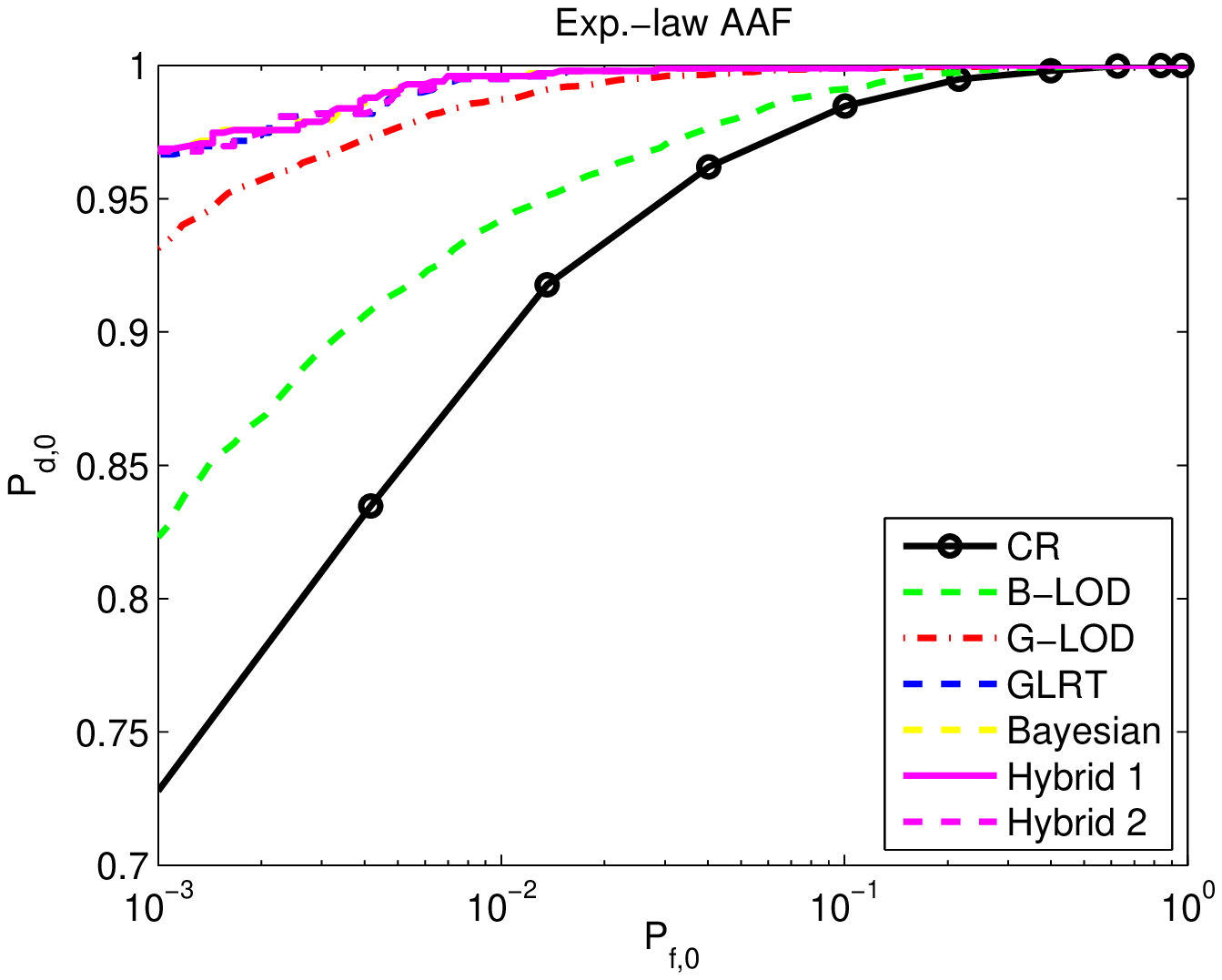}}\caption{$P_{d,0}$ vs. $P_{f,0}$ for all the presented rules (informative
prior setup); WSN with $K=64$ sensors, $\mathrm{SNR}=10\,\mathrm{dB}$,
$P_{e,k}=0$ (ideal reporting channels).\label{fig: ROC ideal BEP K =00003D 64 (informative setup)}}
\end{figure}

Then, in Figs. \ref{fig: ROC ideal BEP K =00003D 64} and \ref{fig: ROC ideal BEP K =00003D 64 (informative setup)}
we illustrate performance for the previous two setups (common and
informative setups) in the case of $K=64$ sensors (i.e. a more densely
deployed WSN), arranged as shown in Fig. \ref{fig: Regular WSN deployment}
(red ``$\circ$'' markers). It is apparent that all rules benefit
from an increase of the number of sensors. Nonetheless, analogous
trends as the case $K=49$ can be observed. 
 
\begin{figure}
\centering{}\subfloat[Power-law AAF.]{\centering{}\includegraphics[width=0.42\paperwidth]{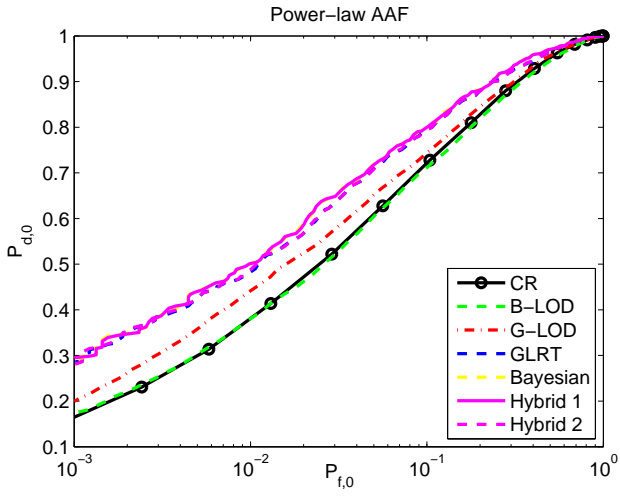}}\subfloat[Exponential AAF.]{\centering{}\includegraphics[width=0.42\paperwidth]{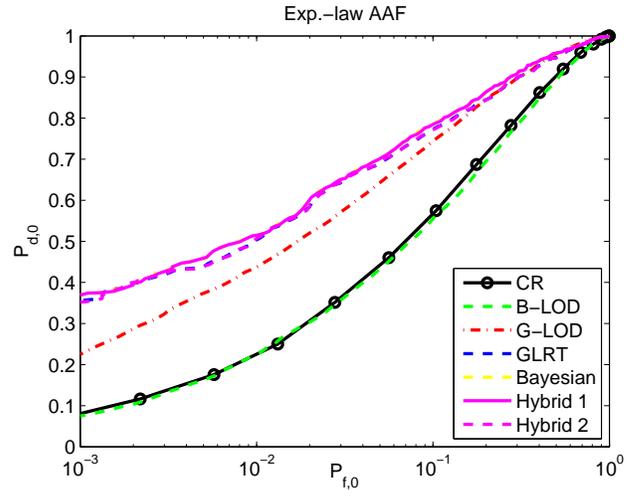}}\caption{$P_{d,0}$ vs. $P_{f,0}$ for all the presented rules; WSN with $K=64$
sensors, $\mathrm{SNR}=10\,\mathrm{dB}$, $P_{e,k}=0.1$ (imperfect
reporting channels).\label{fig: ROC nonideal BEP K =00003D 64} }
\end{figure}

\begin{figure}
\centering{}\subfloat[Power-law AAF.]{\centering{}\includegraphics[width=0.42\paperwidth]{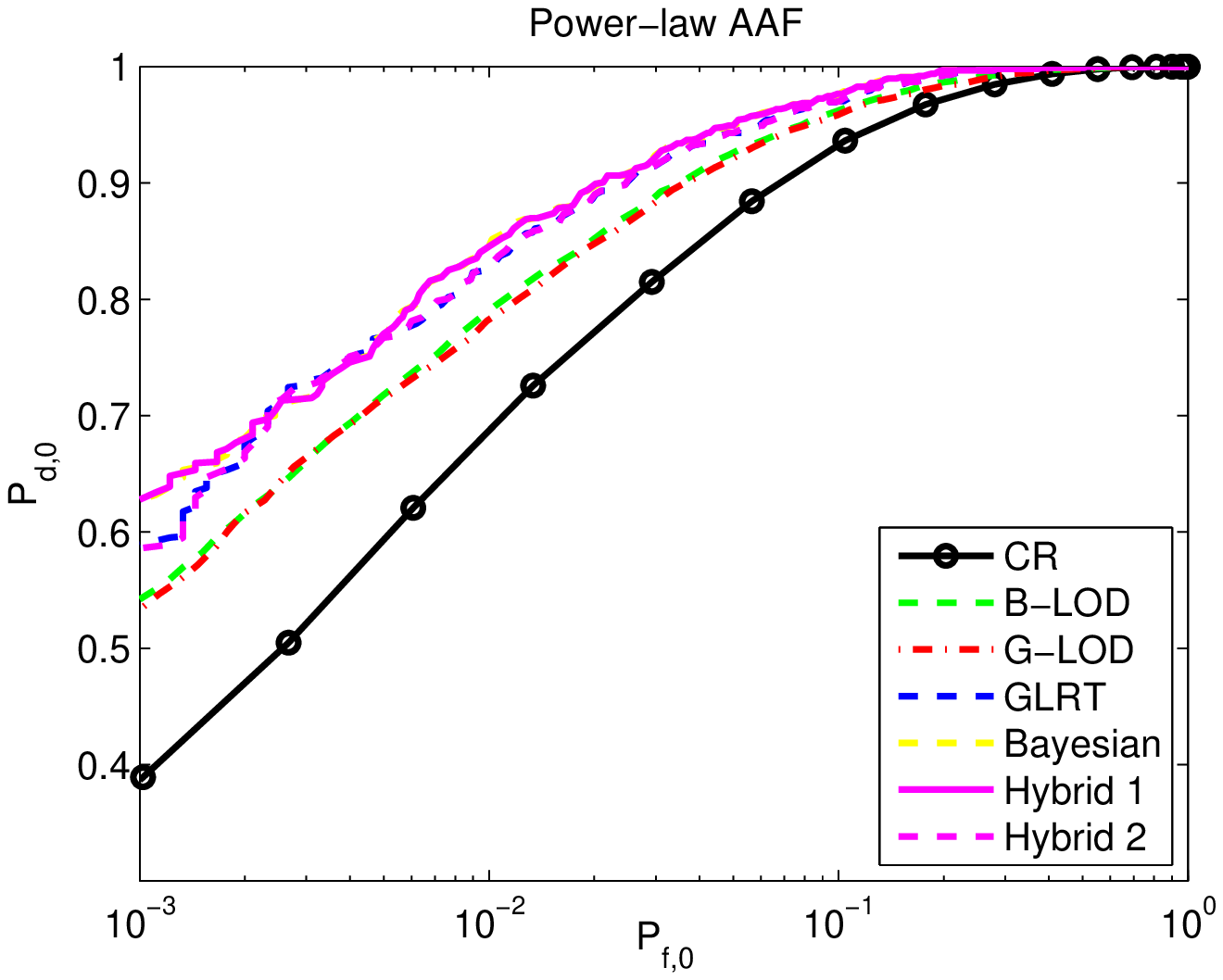}}\subfloat[Exponential AAF.]{\centering{}\includegraphics[width=0.42\paperwidth]{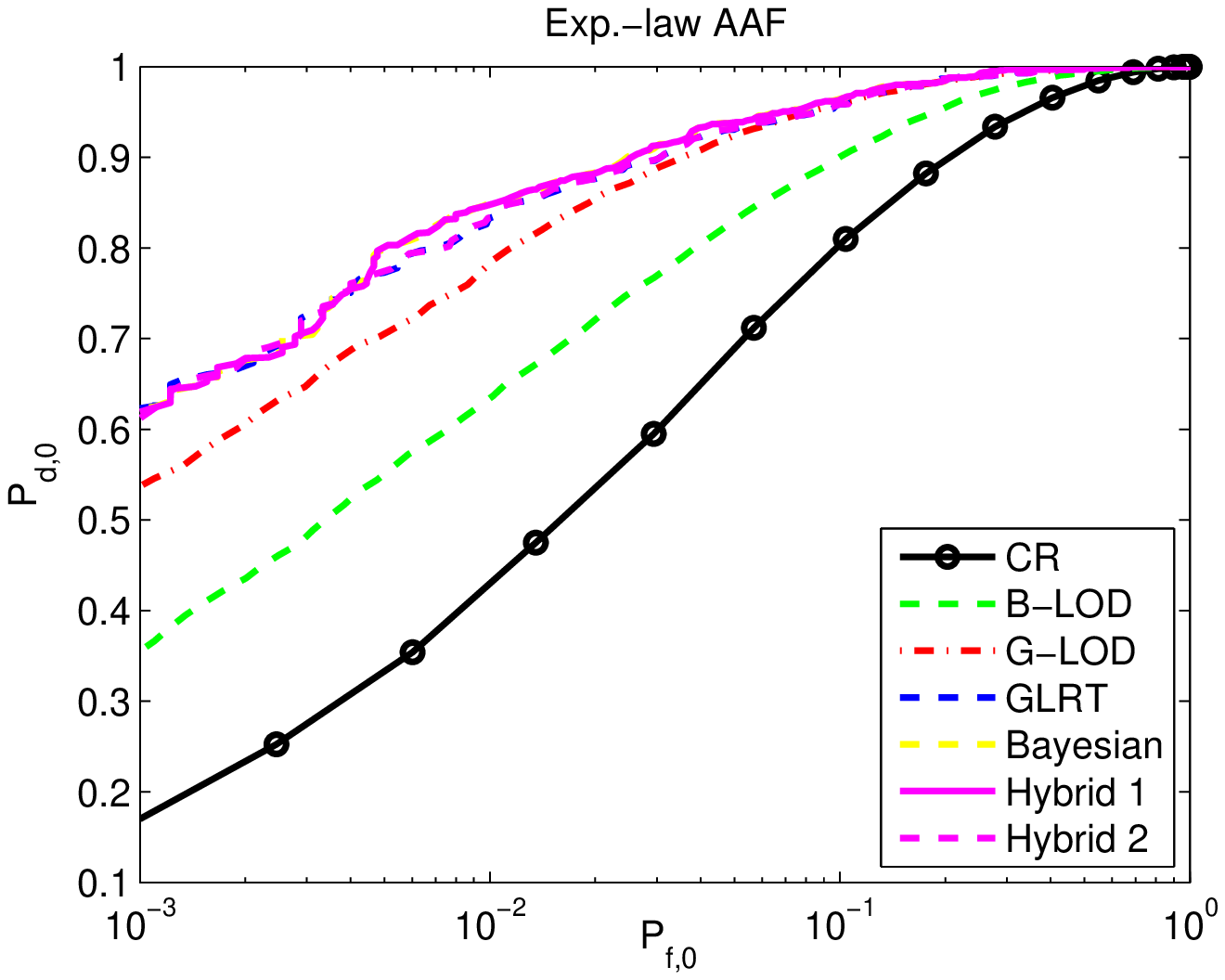}}\caption{$P_{d,0}$ vs. $P_{f,0}$ for all the presented rules (informative
prior setup); WSN with $K=64$ sensors, $\mathrm{SNR}=10\,\mathrm{dB}$,
$P_{e,k}=0.1$ (imperfect reporting channels).\label{fig: ROC nonideal BEP K =00003D 64 (informative setup)} }
\end{figure}

Finally, in Figs. \ref{fig: ROC nonideal BEP K =00003D 64} and \ref{fig: ROC nonideal BEP K =00003D 64 (informative setup)}
we show ROCs for the previous setups (assuming $K=64$) in the case
of imperfect reporting channels. More specifically, for simplicity
we assume the same BEP for all the sensors, i.e. $P_{e,k}=P_{e}$,
$k\in\mathcal{K}$, and we set $P_{e}=0.1$. From both figures it
is apparent a general degradation of performance due to imperfect
reporting channels. Additionally, it can be observed a general decrease
of the performance \emph{spread} for the considered rules. The reason
is that a non-zero BEP tends to \emph{smooth} the spatial signature
of the AAF. Additionally, the equal BEP assumption leads to a similar
relative confidence of each sensor local decision. Therefore the relative
performance loss incurred by CR decreases.

Finally, in Tabs. \ref{tab: Detection rate tab} and \ref{tab: Detection rate tab informative},
we report a comparison of all the presented rules in terms of $P_{d,0}$
(assuming $K=64$ and $\mathrm{SNR=10\,\mathrm{dB}}$) for the relevant
scenario of $P_{f,0}=10^{-2}$, in both the previously considered
cases of uninformative and informative prior ($p(\bm{x}_{T})$), respectively.
As an example, in Tab. \ref{tab: Detection rate tab} when $P_{e,k}=0$
and the assumed AAF follows the exponential law, G-LOD is able to
provide a $26\%$ improvement of the detection rate with respect to
B-LOD and CR.

\begin{table*}
\begin{centering}
\begin{tabular}{c|c|c|c|c}
\hline 
\multirow{2}{*}{\textbf{Fusion Rule}} & \textbf{Pow-Law,} & \textbf{Exp-Law,} & \textbf{Pow-Law, } & \textbf{Exp-Law,}\tabularnewline
 & \textbf{ $P_{e,k}=0$ } & \textbf{ $P_{e,k}=0$} & \textbf{$P_{e,k}=0.1$} & \textbf{ $P_{e,k}=0.1$}\tabularnewline
\hline 
\hline 
$\Lambda_{\mathrm{G}}$ & $0.87$ & $0.83$ & $0.49$ & $0.50$\tabularnewline
\hline 
$\Lambda_{\mathrm{B}}$ & $0.87$ & $0.83$ & $0.5$ & $0.51$\tabularnewline
\hline 
$\Lambda_{\mathrm{GB1}}$ & $0.87$ & $0.83$ & $0.5$ & $0.51$\tabularnewline
\hline 
$\Lambda_{\mathrm{GB2}}$ & $0.87$ & $0.83$ & $0.49$ & $0.50$\tabularnewline
\hline 
$\Lambda_{{\scriptscriptstyle \mathrm{BLOD}}}$ & $0.75$ & $0.55$ & $0.38$ & $0.23$\tabularnewline
\hline 
$\Lambda_{{\scriptscriptstyle \mathrm{CR}}}$ & $0.77$ & $0.55$ & $0.38$ & $0.23$\tabularnewline
\hline 
$\Lambda_{{\scriptscriptstyle \mathrm{GLOD}}}$ & $0.81$ & $0.81$ & $0.44$ & $0.44$\tabularnewline
\hline 
\end{tabular}
\par\end{centering}

\caption{Global detection probability ($P_{d,0})$ comparison for the presented
rules (global false-alarm rate is set to $P_{f,0}=10^{-2}$). WSN
with $K=64$ sensors, $\mathrm{SNR}=10\,\mathrm{dB}$. \label{tab: Detection rate tab}}
\end{table*}
\begin{table*}
\begin{centering}
\begin{tabular}{c|c|c|c|c}
\hline 
\multirow{2}{*}{\textbf{Fusion Rule}} & \textbf{Pow-Law,} & \textbf{Exp-Law,} & \textbf{Pow-Law, } & \textbf{Exp-Law,}\tabularnewline
 & \textbf{ $P_{e,k}=0$ } & \textbf{ $P_{e,k}=0$} & \textbf{$P_{e,k}=0.1$} & \textbf{ $P_{e,k}=0.1$}\tabularnewline
\hline 
\hline 
$\Lambda_{\mathrm{G}}$ & $0.99$ & $0.99$ & $0.83$ & $0.83$\tabularnewline
\hline 
$\Lambda_{\mathrm{B}}$ & $0.99$ & $0.99$ & $0.84$ & $0.85$\tabularnewline
\hline 
$\Lambda_{\mathrm{GB1}}$ & $0.99$ & $0.99$ & $0.83$ & $0.85$\tabularnewline
\hline 
$\Lambda_{\mathrm{GB2}}$ & $0.99$ & $0.99$ & $0.84$ & $0.83$\tabularnewline
\hline 
$\Lambda_{{\scriptscriptstyle \mathrm{BLOD}}}$ & $0.98$ & $0.94$ & $0.78$ & $0.64$\tabularnewline
\hline 
$\Lambda_{{\scriptscriptstyle \mathrm{CR}}}$ & $0.97$ & $0.90$ & $0.72$ & $0.47$\tabularnewline
\hline 
$\Lambda_{{\scriptscriptstyle \mathrm{GLOD}}}$ & $0.98$ & $0.98$ & $0.78$ & $0.78$\tabularnewline
\hline 
\end{tabular}
\par\end{centering}

\caption{Global detection probability ($P_{d,0})$ comparison for the presented
rules (global false-alarm rate is set to $P_{f,0}=10^{-2}$). WSN
with $K=64$ sensors (informative prior setup), $\mathrm{SNR}=10\,\mathrm{dB}$.
\label{tab: Detection rate tab informative}}
\end{table*}

\section{Conclusions and Future Directions\label{sec: Conclusions}}

In this paper we tackled distributed detection of a non-cooperative
target. Sensors measure an unknown random signal (embedded in Gaussian
noise) with an AAF depending on the distance between the sensor and
the target (unknown) positions. Each local decision, based on (local)
energy detection, is then sent to a FC for improved detection performance.
The focus of this work has concerned the development of practical
fusion rules at the FC. To this end, we first focused on the scenario
where the emitted power is available at the FC and analyzed fusion
rules based on GLRT and Bayesian approaches. Then we moved to the
more realistic case of unknown target location ($\bm{x}_{T}$) and
power ($\sigma_{s}^{2}$). Such case is typical when detecting non-cooperative
targets. The present problem has been formally cast as a one-sided
hypothesis testing with nuisance parameters (i.e. $\bm{x}_{T}$) which
are present only under $\mathcal{H}_{1}$ (viz. target-present hypothesis).
For the resulting hypothesis testing, we analyzed several fusion rules
based on: ($i)$ GLRT, ($ii$) Bayesian approach and ($iii$) hybrid
combinations of the two. All these rules have been shown to achieve
similar performance in all the scenarios being considered. Unfortunately,
they all require a grid-based implementation on the Cartesian product
of optimization (integration) space of $\bm{x}_{T}$ and $\sigma_{s}^{2}$.
Then, with the intent of reducing the computational complexity required
by all these approaches, we proposed other two (sub-optimal) fusion
rules built upon the LOD framework \cite{Kassam1988} and based on
the following specific rationales:
\begin{itemize}
\item B-LOD: $\bm{x}_{T}$ is treated as a random parameter with a prior
pdf $p(\bm{x}_{T})$, while $\sigma_{s}^{2}$ is tackled under the
LOD framework;
\item G-LOD: A generalized version of LOD (based on \cite{Davies1987}),
arising from maximization (w.r.t. nuisance parameter $\bm{x}_{T}$)
of a family of LOD decision statistics obtained by assuming $\bm{x}_{T}$
known;
\end{itemize}
The aforementioned rules present reduced complexity with respect to
the previous rules, all requiring a grid-based search or integration
with respect to both $\bm{x}_{T}$ and $\sigma_{s}^{2}$. More specifically,
B-LOD retains a linear complexity in the number of sensors (as the
simple CR), while G-LOD is based on a (reduced) grid search which
only requires optimization w.r.t. $\bm{x}_{T}$. Additionally, G-LOD
has been shown to outperform CR in all the considered cases and to
incur in a moderate performance loss with respect to other rules requiring
grid implementation on both parameters. Differently, B-LOD has been
shown to provide a significant gain over CR only when the prior pdf
of $\bm{x}_{T}$ is informative enough.

All the considered rules have been extended to the case of imperfect
reporting channels, modeled as BSCs with corresponding BEPs $P_{e,k}$
assumed known at the FC. It has been demonstrated that only a slight
modification of their expressions is required in order to account
for this additional uncertainty, whereas it has been observed that
non-zero BEPs tend to smooth the spatial signature determined by the
AAF and thus to reduce the gain obtained by all the rules exploiting
spatial information of the target w.r.t. CR.

Future works will include the design and analysis of fusion rules
based on soft-decisions (i.e. multi-bit quantization) from the sensors,
as well as the problem of detecting time-evolving (diffusive) sources
with possibly moving sensors. Both the cases of cooperative and uncooperative
targets are of interest. Furthermore, the case of uncertain sensors
positions will be tackled in comparison to the well-known concept
of scan statistics \cite{Guerriero2009}. Finally, robust design of
fusion rules accounting for uncertainties at the reporting channels
(i.e. unknown BEPs) will be also considered.

\section*{Appendix}

\subsection*{\noindent Derivation of Bayesian LOD}

\noindent In this Appendix, we derive the explicit expression of the
LOD \cite{Kassam1988} based on a prior distribution assumption for the
target position $\bm{x}_{T}$, that is based on Eq. (\ref{eq: pdf H1 Bayesian LOD}).
To this end, starting from the implicit form in (\ref{eq: Bayesian LOD - general (BSC)}),
we first concentrate on obtaining the closed form of $\frac{\partial\ln\left[P(\bm{d}|\mathcal{H}_{1};\theta)\right]}{\partial\theta}$.
The latter term is obtained recalling that:
\begin{equation}
\ln\left[P(\bm{d}|\mathcal{H}_{1};\theta)\right]=\ln\left[\int\prod_{k=1}^{K}P_{d,k}(\bm{x}_{T},\theta)^{d_{k}}\left[1-P_{d,k}(\bm{x}_{T},\theta)\right]^{(1-d_{k})}\,p(\bm{x}_{T})\,d\bm{x}_{T}\right]\,.
\end{equation}
The derivative of the log-pdf can be thus obtained in closed form
as:
\begin{equation}
\frac{\partial\ln\left[P(\bm{d}|\mathcal{H}_{1};\theta)\right]}{\partial\theta}=\frac{\int\frac{\partial P(\bm{d}|\mathcal{H}_{1};\bm{x}_{T},\theta)}{\partial\theta}\,p(\bm{x}_{T})\,d\bm{x}_{T}}{\int\prod_{k=1}^{K}P_{d,k}(\bm{x}_{T},\theta)^{d_{k}}\left[1-P_{d,k}(\bm{x}_{T},\theta)\right]^{(1-d_{k})}\,p(\bm{x}_{T})\,d\bm{x}_{T}}\,,\label{eq: BayLod gradlogpdf}
\end{equation}
where we have \emph{interchanged} the order of derivative and integration
at the numerator. Also, the derivative within the integral in (\ref{eq: BayLod gradlogpdf})
at the numerator can be evaluated in explicit form as:
\begin{align}
\frac{\partial P(\bm{d}|\mathcal{H}_{1};\bm{x}_{T},\theta)}{\partial\theta}= & \left(\prod_{k=1}^{K}P_{d,k}(\bm{x}_{T},\theta)^{d_{k}}\left[1-P_{d,k}(\bm{x}_{T},\theta)\right]^{(1-d_{k})}\right)\,\nonumber \\
 & \times\sum_{k=1}^{K}\frac{d_{k}-P_{d,k}(\bm{x}_{T},\theta)}{P_{d,k}(\bm{x}_{T},\theta)\left[1-P_{d,k}(\bm{x}_{T},\theta)\right]}\frac{\partial P_{d,k}(\bm{x}_{T},\theta)}{\partial\theta}\,.\label{eq: Prob derivative}
\end{align}
For the considered model in Eq. (\ref{eq: Pd Pfa local sensors}),
the derivative of the $P_{d,k}$ w.r.t. $\theta$ is given explicitly
as:
\begin{align}
\frac{\partial P_{d,k}(\bm{x}_{T},\theta)}{\partial\theta} & =2\,\frac{\partial}{\partial\theta}\mathcal{Q}\left(\sqrt{\frac{\gamma_{k}}{\sigma_{w,k}^{2}+\theta\,g^{2}(\bm{x}_{T},\bm{x}_{k})}}\right)\\
 & =p_{w}\left(\sqrt{\frac{\gamma_{k}}{\sigma_{w,k}^{2}+\theta\,g^{2}(\bm{x}_{T},\bm{x}_{k})}}\right)\frac{\sqrt{\gamma_{k}}\,g^{2}(\bm{x}_{T},\bm{x}_{k})}{\left[\sigma_{w,k}^{2}+\theta\,g^{2}(\bm{x}_{T},\bm{x}_{k})\right]^{3/2}}\,.\label{eq: Det prob derivative Bayesian LOD}
\end{align}
Evaluating the derivative of the log-pdf in (\ref{eq: BayLod gradlogpdf})
at $\theta=\theta_{0}$ (which corresponds to null hypothesis $\mathcal{H}_{0}$,
see Eq. (\ref{eq: Hypothesis testing -  Unknown Target Power})),
leads to
\begin{equation}
\left.\frac{\partial\ln\left[P(\bm{d}|\mathcal{H}_{1};\theta)\right]}{\partial\theta}\right|_{\theta=\theta_{0}}=\frac{\int\left.\frac{\partial P(\bm{d};\mathcal{H}_{1},\bm{x}_{T},\theta)}{\partial\theta}\right|_{\theta=0}\,p(\bm{x}_{T})\,d\bm{x}_{T}}{\prod_{k=1}^{K}(P_{f,k})^{d_{k}}(1-P_{f,k})^{(1-d_{k})}}\,,
\end{equation}
where, exploiting (\ref{eq: Prob derivative}), we obtain: 
\begin{equation}
\left.\frac{\partial P(\bm{d}|\mathcal{H}_{1};\bm{x}_{T},\theta)}{\partial\theta}\right|_{\theta=\theta_{0}}=\left(\prod_{k=1}^{K}(P_{f,k})^{d_{k}}\left[1-P_{f,k}\right]^{(1-d_{k})}\right)\sum_{k=1}^{K}\frac{d_{k}-P_{f,k}}{P_{f,k}\left(1-P_{f,k}\right)}\left.\frac{\partial P_{d,k}(\bm{x}_{T},\theta)}{\partial\theta}\right|_{\theta=\theta_{0}}\,,
\end{equation}
and in turn (cf. Eq. (\ref{eq: Det prob derivative Bayesian LOD}))
\begin{equation}
\left.\frac{\partial P_{d,k}(\bm{x}_{T},\theta)}{\partial\theta}\right|_{\theta=\theta_{0}}=p_{w}\left(\sqrt{\frac{\gamma_{k}}{\sigma_{w,k}^{2}}}\right)\frac{\sqrt{\gamma_{k}}\,g^{2}(\bm{x}_{T},\bm{x}_{k})}{\left(\sigma_{w,k}^{2}\right)^{3/2}}\,.
\end{equation}
Then, exploiting the appropriate substitutions, we obtain:
\begin{equation}
\left.\frac{\partial\ln\left[P(\bm{d}|\mathcal{H}_{1};\theta)\right]}{\partial\theta}\right|_{\theta=\theta_{0}}=\sum_{k=1}^{K}\frac{d_{k}-P_{f,k}}{P_{f,k}\left(1-P_{f,k}\right)}\,p_{w}\left(\sqrt{\frac{\gamma_{k}}{\sigma_{w,k}^{2}}}\right)\frac{\sqrt{\gamma_{k}}}{\left(\sigma_{w,k}^{2}\right)^{3/2}}\left(\int g^{2}(\bm{x}_{T},\bm{x}_{k})\,p(\bm{x}_{T})\,d\bm{x}_{T}\right)\,.\label{eq: gradient log pdf Bayesian}
\end{equation}
Now we show how to obtain the explicit form of the FI evaluated at
$\theta_{0}$. First, we start from the common definition: 
\begin{equation}
I(\theta_{0})=\mathbb{E}_{P(\bm{d}|\mathcal{H}_{0})}\left\{ \left(\left.\frac{\partial\ln\left[P(\bm{d}|\,\mathcal{H}_{1};\theta)\right]}{\partial\theta}\right|_{\theta=\theta_{0}}\right)^{2}\right\} \,.
\end{equation}
Since the elements $d_{k}$, $k=1,\ldots K$, are uncorrelated it
holds that:
\begin{equation}
I(\theta_{0})=\sum_{k=1}^{K}\frac{\mathbb{E}\left\{ \left(d_{k}-P_{f,k}\right)^{2}\right\} }{P_{f,k}^{2}\left(1-P_{f,k}\right)^{2}}\,p_{w}^{2}\left(\sqrt{\frac{\gamma_{k}}{\sigma_{w,k}^{2}}}\right)\frac{\gamma_{k}}{\left(\sigma_{w,k}^{2}\right)^{3}}\left(\int g^{2}(\bm{x}_{T},\bm{x}_{k})\,p(\bm{x}_{T})\,d\bm{x}_{T}\right)^{2}\,.
\end{equation}
Evaluating the expectation inside the above equation, provides the
explicit form of $I(\theta_{0})$:
\begin{equation}
I(\theta_{0})=\sum_{k=1}^{K}\frac{1}{P_{f,k}\left(1-P_{f,k}\right)}\,p_{w}^{2}\left(\sqrt{\frac{\gamma_{k}}{\sigma_{w,k}^{2}}}\right)\frac{\gamma_{k}}{\left(\sigma_{w,k}^{2}\right)^{3}}\left(\int g^{2}(\bm{x}_{T},\bm{x}_{k})\,p(\bm{x}_{T})\,d\bm{x}_{T}\right)^{2}\,.\label{eq: Bayesian Fisher closed form}
\end{equation}
Substitution of closed forms of Eqs. (\ref{eq: gradient log pdf Bayesian})
and (\ref{eq: Bayesian Fisher closed form}) into the implicit form
in (\ref{eq: Bayesian LOD - general}), provides the explicit expression
reported in Eq. (\ref{eq: Bayesian LOD - explicit}).

\subsection*{\noindent Derivation of Generalized LOD}

In this Appendix, we derive the explicit expression of the G-LOD proposed
by Davies. To this end, we concentrate on finding the explicit form
of LOD fusion rule \cite{Kassam1988} assuming $\bm{x}_{T}$ known.
Once obtained, the explicit expression will clearly depend on $\bm{x}_{T}$.
Such expression will be then plugged in the maximization of Eq. (\ref{eq: Davies LOD general})
to obtain the final statistic. First, we observe that:
\begin{equation}
\frac{\partial\ln P(\bm{d}|\mathcal{H}_{1};\bm{x}_{T},\theta)}{\partial\theta}=\sum_{k=1}^{K}\frac{d_{k}-P_{d,k}(\bm{x}_{T},\theta)}{P_{d,k}(\bm{x}_{T},\theta)\left[1-P_{d,k}(\bm{x}_{T},\theta)\right]}\frac{\partial P_{d,k}(\bm{x}_{T},\theta)}{\partial\theta}\,,
\end{equation}
where $\frac{\partial P_{d,k}(\bm{x}_{T},\theta)}{\partial\theta}$
is defined as in Eq. (\ref{eq: Det prob derivative Bayesian LOD}).
Setting $\theta=\theta_{0}$, the above term reduces to:
\begin{equation}
\left.\frac{\partial\ln P(\bm{d}|\mathcal{H}_{1};\bm{x}_{T},\theta)}{\partial\theta}\right|_{\theta=\theta_{0}}=\sum_{k=1}^{K}\frac{d_{k}-P_{f,k}}{P_{f,k}(1-P_{f,k})}\,p_{w}\left(\sqrt{\frac{\gamma_{k}}{\sigma_{w,k}^{2}}}\right)\frac{\sqrt{\gamma_{k}}\,g^{2}(\bm{x}_{T},\bm{x}_{k})}{\left(\sigma_{w,k}^{2}\right)^{3/2}}\,,\label{eq: grad log-pdf Davies theta_0}
\end{equation}
where we exploited the definition in Eq. (\ref{eq: Det prob derivative Bayesian LOD}).
Similarly, exploiting (conditional) independence of the decisions
$d_{k}$, $k=1,\ldots K$, we obtain:
\begin{equation}
I(\bm{x}_{T},\theta)=\sum_{k=1}^{K}I_{k}(\bm{x}_{T},\theta)\,,
\end{equation}
where we have denoted with $I_{k}(\bm{x}_{T},\theta)$ the contribution
of $k$th to the FI, that is:
\begin{align}
I_{k}(\bm{x}_{T},\theta) & =\mathbb{E}\left\{ \left(\frac{\partial\ln\left[P(d_{k}|\mathcal{H}_{1};\bm{x}_{T},\theta)\right]}{\partial\theta}\right)^{2}\right\} \\
 & =\frac{\mathbb{E}\left\{ \left[d_{k}-P_{d,k}(\bm{x}_{T},\theta)\right]^{2}\right\} }{P_{d,k}(\bm{x}_{T},\theta)^{2}\left[1-P_{d,k}(\bm{x}_{T},\theta)\right]^{2}}\left(\frac{\partial P_{d,k}(\bm{x}_{T},\theta)}{\partial\theta}\right)^{2}\label{eq: MLOD - Kth FI - snd line}\\
 & =\frac{1}{P_{d,k}(\bm{x}_{T},\theta)\left[1-P_{d,k}(\bm{x}_{T},\theta)\right]}\left(\frac{\partial P_{d,k}(\bm{x}_{T},\theta)}{\partial\theta}\right)^{2}
\end{align}
where, in obtaining the last line we have explicitly evaluated the
expectation in Eq. (\ref{eq: MLOD - Kth FI - snd line}). Then, substitution
$\theta\rightarrow\theta_{0}$ in $I(\bm{x}_{T},\theta)$ provides:
\begin{align}
I(\bm{x}_{T},\theta_{0})\;= & \sum_{k=1}^{K}I_{k}(\bm{x}_{T},\theta_{0})\\
= & \sum_{k=1}^{K}\frac{1}{P_{f,k}\left[1-P_{f,k}\right]}p_{w}^{2}\left(\sqrt{\frac{\gamma_{k}}{\sigma_{w,k}^{2}}}\right)\frac{\gamma_{k}\,g^{4}(\bm{x}_{T},\bm{x}_{k})}{\left(\sigma_{w,k}^{2}\right)^{3}}\,.\label{eq: MLOD - Kth FI - theta_0}
\end{align}
Exploiting Eqs. (\ref{eq: MLOD - Kth FI - theta_0}) and (\ref{eq: grad log-pdf Davies theta_0})
into (\ref{eq: Davies LOD general}), provides the final expression
in (\ref{eq: Davies LOD final}).

\section*{References}

\bibliographystyle{elsarticle-num}
\bibliography{sensor_networks}

\end{document}